\documentclass[sigconf,screen]{acmart}
\PassOptionsToPackage{prologue,dvipsnames}{xcolor}
\copyrightyear{2024}
\acmYear{2024}
\setcopyright{rightsretained}
\acmConference[CCS '24] {Proceedings of the 2024 ACM SIGSAC Conference on Computer and Communications Security}{October 14--18, 2024}{Salt Lake City, UT, USA.}
\acmBooktitle{Proceedings of the 2024 ACM SIGSAC Conference on Computer and Communications Security (CCS '24), October 14--18, 2024, Salt Lake City, UT, USA}
\acmISBN{979-8-4007-0636-3/24/10}
\acmDOI{10.1145/3658644.3690188}

\usepackage{etoolbox}
\makeatletter
\patchcmd{\maketitle}{\@copyrightpermission}{
  \begin{minipage}{0.3\columnwidth}
    \href{https://creativecommons.org/licenses/by-nc-sa/4.0/}{
    \includegraphics[width=0.90\textwidth]{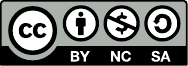}
    }
  \end{minipage}\hfill
  \begin{minipage}{0.7\columnwidth}
    \href{https://creativecommons.org/licenses/by-nc-sa/4.0/}{This work is licensed under a Creative Commons Attribution-NonCommercial-ShareAlike International 4.0 License.}
  \end{minipage}
}{}{}
\makeatother

\makeatletter
\patchcmd{\maketitle}{\@copyrightpermission}{
  \begin{minipage}{0.3\columnwidth}
    \href{https://creativecommons.org/licenses/by-nc-sa/4.0/}{\includegraphics[width=0.90\textwidth]{fig/by-nc-sa.pdf}}
  \end{minipage}\hfill
  \begin{minipage}{0.7\columnwidth}
    \href{https://creativecommons.org/licenses/by-nc-sa/4.0/}{This work is licensed under a Creative Commons Attribution-NonCommercial-ShareAlike International 4.0 License.}
  \end{minipage}
}{}{}
\makeatother

\usepackage[normalem]{ulem}

\usepackage{nameref}

\usepackage{endnotes,xspace,graphicx,fancyvrb,multirow}
\usepackage{microtype}
\usepackage{booktabs}
\usepackage{colortbl}
\usepackage[T1]{fontenc}
\usepackage[utf8]{inputenc}
\usepackage{fancyhdr}
\usepackage{enumitem}

\usepackage{fp}
\usepackage[binary-units,mode=text]{siunitx}
\usepackage{balance}

\usepackage{bm}

\usepackage[toc,acronym,nowarn]{glossaries}
\glsdisablehyper

\definecolor{SunsetOrange}{RGB}{255,96,92}
\definecolor{PastelOrange}{RGB}{255,189,68}
\definecolor{Malachite}{RGB}{0,202,78}

\definecolor{MSYello}{RGB}{255, 187, 0}
\definecolor{MSGreen}{RGB}{124, 187, 0}
\definecolor{MSBlue}{RGB}{0, 161, 241}
\definecolor{MSRed}{RGB}{246, 83, 20}

\definecolor{mydarkblue}{rgb}{0.0,0.0,0.5}
\definecolor{myblue}{rgb}{0.0,0.0,1.0}
\definecolor{pink}{rgb}{1.0,0.1,0.8}

\usepackage[caption=false, font=footnotesize]{subfig}
\usepackage{tikz}
\usepackage{xspace}
\usepackage{booktabs}
\usepackage{enumitem}
\usepackage{colortbl}
\usepackage{adjustbox}

\usepackage{multirow} 
\usepackage{balance}
\usepackage{stfloats}
\usepackage{placeins}
\usepackage[utf8]{inputenc}
\usepackage{amsmath}
\usepackage{tikz}
\usepackage{xcolor}
\usepackage{amsfonts}
\usepackage{hyperref}
\AtBeginDocument{

\hypersetup{
  colorlinks,
  citecolor=blue,                    
  linkcolor=blue,                         
}
}

\usepackage{cleveref}
\usepackage{pifont}
\crefformat{section}{#2\S{}#1#3}
\crefformat{subsection}{#2\S{}#1#3}  
\Crefformat{subsection}{#2\S{}#1#3}  
\crefformat{figure}{#2Fig.~#1#3}  
\Crefformat{figure}{#2Fig.~#1#3}
\crefformat{subfigure}{#2Fig.~#1#3}
\Crefformat{subfigure}{#2Fig.~#1#3}
\crefformat{table}{#2Tab.~#1#3}
\Crefformat{table}{#2Tab.~#1#3}
\usepackage{threeparttable}

\crefname{section}{§}{§§}
\Crefname{section}{§}{§§}
\usepackage{url}

\usepackage[]{appendix}
\usepackage{textcomp}
\usepackage{xurl}
\usepackage{tcolorbox}

\usepackage{shortcuts}

\definecolor{patriarch}{rgb}{0.5, 0.0, 0.5}

\definecolor{citecolor}{RGB}{7, 48, 88}
\definecolor{linkcolor}{RGB}{162, 0, 2}
\definecolor{urlcolor}{RGB}{7, 58, 110}
\definecolor{acmblue}{cmyk}{1, 0.1, 0, 0.1}
\definecolor{acmdarkblue}{cmyk}{1, 0.2, 0, 0.15}

\begin{document}
\title{The \codename's Guide to High-Assurance System Observability Protection with Efficient Permission Switches}

\author{Chuqi Zhang}
\authornote{\footnotesize{Work partly completed while being a visiting researcher at ASU with Dr. Adil Ahmad. }}
\affiliation{%
  \country{}
  \department{\small{School of Computing}}
  \institution{\small{National University of Singapore}}
}

\author{Jun Zeng}
\affiliation{%
  \country{}
  \institution{\small{Independent Researcher}}
}

\author{Yiming Zhang}
\affiliation{%
  \country{}
  \institution{\small{Southern University of Science and Technology}}
}
\affiliation{%
  \country{}
  \institution{\small{The Hong Kong Polytechnic University}}
}

\author{Adil Ahmad}
\authornote{Corresponding author.}
\affiliation{%
  \country{}
  \department{\small{School of Computing and Augmented Intelligence}}
  \institution{\small{Arizona State University}}
}

\author{Fengwei Zhang}
\affiliation{%
  \country{}
  \department{\small{Department of Computer Science and Engineering}}
  \institution{\small{Southern University of Science and Technology}}
}

\author{Hai Jin}
\affiliation{%
  \country{}
  \institution{\small{Huazhong University of Science and Technology}}
}

\author{Zhenkai Liang}
\affiliation{%
  \country{}
  \department{\small{School of Computing}}
  \institution{\small{National University of Singapore}}
}
\renewcommand{\shortauthors}{Chuqi Zhang et al.}

\thispagestyle{empty}

\begin{abstract}

Protecting system observability records ({\em logs}) from compromised OSs has gained significant traction in recent times, with several note-worthy approaches proposed.
Unfortunately, none of the proposed approaches achieve high performance with tiny log protection delays. 
They also leverage risky environments for protection (\eg many use general-purpose hypervisors or TrustZone, which have large TCB and attack surfaces).
%
\codename is an attempt to rectify this problem.
The system is designed to ensure (a) in-memory protection of batched logs within a short and configurable real-time deadline by efficient hardware permission switching, and (b) an end-to-end high-assurance environment built upon hardware protection primitives with debloating strategies for secure log protection, persistence, and management.
%
Security evaluations and validations show that \codename reduces log protection delay by 93.3--99.3\% compared to the state-of-the-art, while reducing TCB by 9.4--26.9$\times$.
Performance evaluations show \codename incurs a geometric mean of less than 6\% overhead on diverse real-world programs, improving on the state-of-the-art approach by 61.9--77.5\%.

\end{abstract}



\maketitle

\section{Introduction}
\label{sec:intro}

System observability is the capability achieved by {\em logging} sensitive activities across {different system layers}, including the operating system~(OS)~\cite{king2003backtracking}, network~\cite{zhou2011securenet}, and applications~\cite{hassan2020omegalog}. 
In modern enterprises, logs provide invaluable insights into attacks for {post-mortem} investigations: a report notes that 75\% of analysts find logs to be the most valuable forensic investigation asset~\cite{vmware_survey}.

Alas, attackers are well aware of the importance of logs, and thus intentionally destroy or tamper with logs to hide their footprints in victim computers~\cite{destroy_survey}.
They typically achieve this after escalating privilege into the OS~\cite{destroy_survey}---a realistic threat given the complex and buggy codebase of commodity kernels~\cite{dmitry2016aslr,lin2022dirtycred}.
%


To counter log tampering and ensure attack investigations, several systems have been proposed to protect logs against a compromised OS.
%
They provide {\em tamper-evident} log hashes~\cite{paccagnella2020custos,riccardo2020kennylogging,hoang2022faster,karande2017sgxlog,shepherd2017emlog}, or preserve both integrity and availability~\cite{dunlap2002revirt,ahmad2022hardlog,varun2023omnilog,ahmad2024veil} of the logs {\em before} system compromise.
While these systems take significant strides towards system observability protection, they still have the following limitations (\cref{sec:motivation}).



\begin{itemize}[leftmargin=*]
\item 
{\em Risky protection environments with large TCB and attack surfaces}.
In-memory protection solutions~\cite{varun2023omnilog,shepherd2017emlog,dunlap2002revirt} employ general-purpose privileged components~(\eg hypervisors with guest virtual machines, or TrustZone with secure OSs) to isolate host memory and manage logs.
These full-fledged components maintain bloated interfaces to support their feature-rich clients (virtual machines or trusted applications), which are not directly required by log protection.
As such, their redundant interfaces bring notorious attack surfaces~\cite{cerdeira2020tzsok,chen2023security,Shi2017DeconstructingXen}.
\item 
{\em Significant synchronization slowdowns in I/O intensive workloads.}
For those in-host-memory protection systems~\cite{varun2023omnilog,shepherd2017emlog,ahmad2024veil}, as soon as each log is generated, it is {\em copied} to the isolated privileged memory to be managed and finally persisted (onto a protected disk).
Considering the high log throughput under realistic workloads within different layers of sources, our evaluation shows that this incurs up to a 52.6\% slowdown, which is considerable given the tight budget of production systems~\cite{kasikci2017lazy}. 

\item
{\em Large exposure protection time windows for log tampering attacks.}
Alternative solutions~\cite{sekar2024eaudit,ahmad2022hardlog,robert2018pesos} then batch logs in the untrusted host memory, and periodically protect logs by transferring them into an external local {\em tamper-proof} device or remote storage.
%
Their inherent I/O latency exposes a large protection window, during which logs remain vulnerable in host memory, with the best case being \textms{15}~\cite{ahmad2022hardlog}.
Our study shows that proof-of-concept (PoC) kernel exploits (not specifically designed for speed) can tamper with {\em all} logs within exposed windows.
\end{itemize}

\codename is an observability protection system built on two key design principles~(\cref{sec:design_overview}) to overcome prior work limitations.

%


To ensure logs are maintained in a high-assurance in-memory environment, \codename takes a first principles approach to environment design~(\cref{subsec:overview_secure_environment}).
In particular, only components directly required for log protection are included, and large software components are {\em debloated} to minimize TCB and attack surface.
Additionally, the task of remote log management is delegated to a {\em protected process} under the native OS.
The combination of these strategies reduces \codename's TCB by $9.4$--$26.9\times$ compared to prior work, and helps build a significantly less vulnerable system.

To achieve high performance while protecting against tampering attacks, \codename ensures that logs are protected in memory asynchronously but within a short, configurable real-time deadline~(\cref{subsec:overview_inmemory_observability_protection}).
In-memory protection is further sped up by using hardware memory permission switching primitives, instead of memory copies.
\codename achieves a protection deadline between \textms{1.015} and \textus{100.12}, 93.3--99.3\% lower than I/O-based asynchronous systems.
Under such settings, the system reduced performance overhead by $61.9-77.5$\% compared to existing work.
%

%

There are three challenges towards \codename's design based on the aforementioned principles (\cref{sec:design}).
First, it is unclear how to design the high-assurance environment without significant hardware or software porting and deployment effort.
Second, it is challenging to {\em enforce} real-time configurable protection deadlines on a system that produces logs at different layers, especially when an attacker may possess unprivileged access before OS compromise.
The third is to natively delegate the log management process execution to an {\em untrusted} OS, with the notion of assured correctness.

\codename leverages platform-available hardware protection primitives~\cite{rme-extension,s2pt,smmu} as the building blocks to maintain its secure environment.
Automatic debloating techniques~\cite{guo2022driverlet,wang2022rttee} are employed to port software components essential for log availability.
Additionally, unmodified native processes with attestation philosophies~\cite{mckeen2013sgx} are adopted, minimizing software redesign efforts (\cref{subsec:design_security_monitor}).
%

%
%
%

To enforce short, configurable protection deadlines, \codename unifies logs produced at different layers into its kernel buffers. 
There, \codename leverages a precise hardware timer for log protection---using real-time memory permission switches---that is isolated from extra delays that a non-privileged attacker may induce.
All protected logs are eventually, yet assuredly, persisted to disk~(\cref{subsec:design_realtime_deadline}).

\codename secures native log management process by memory view and execution context enforcement.
The process's memory permissions are restricted from the untrusted OS, while system calls are secured through transparent context switch interposition.
The integrity of remote tasks is further ensured by end-to-end secure channels and provisioned cryptographic secrets (\cref{subsec:design_daemon}).

We implemented the prototype of \codename on an ARM machine~(\cref{sec:impl}).
Both the Stage-2 Page Table (S2PT~\cite{s2pt}) and the Granule Protection Table (GPT~\cite{rme-extension-website}) are implemented as the memory permission primitive for log protection, respectively.
Observability logs are generated using the extended-Aqua Tracee~\cite{tracee} with eBPF~\cite{ebpf}, 
a widely-used technique for security observability auditing~\cite{agman2021bpfroid, minna2023sok}. 
%
To foster future research, we will release the artifact of our implementation prototype~\cite{hhk-artifact}.

%

Using our prototype, we analyzed and evaluated \codename's log protection guarantees~(\cref{sec:sec_evaluation}).
Worst-case stress testing shows \codename guarantees a \textms{1.015} to \textus{100.13} protection deadline, which is 93.3\% to 99.3\% lower than the state-of-the-art asynchronous protection~\cite{ahmad2022hardlog}.
%
%
Under our most powerful tested attack, \codename protected the vast majority of the log traces ($\ge$97\%) even configured with its most relaxed deadline.
Importantly, all lost logs were for the last step (kernel module loading), while all logs related to exploitation and escalation were always saved.
%


We evaluated \codename's performance~(\cref{sec:performance_evaluation}) with both micro-benchmarks and commonly used real-world programs.
Experimental results show that \codename introduces a geometric mean overhead of 1.8\% for log-sparse and 9.9\% for log-intensive programs (6\% cumulative geometric mean).
The overhead is up to 77.5\% and 61.9\% lower than the state-of-the-art protection system under log-sparse and intensive programs, respectively.
This shows that \codename can be readily deployed today to achieve high performance and secure log protection in enterprise computers.
%






\section{Background}

\subsection{System Observability}
\label{subsec:system_observability}

%
Observability logs provide visibility into enterprise computer's {\em security posture}, and are captured by Security Information and Event Management (SIEM) from the industry~\cite{vmware_survey,SIEM-logs} and academia~\cite{yu2021alchemist,datta2022alastor}.
Based on standard security practices~\cite{obs_sec}, we consider three main sources (or layers) of logs: (a) {\em application logs} for program execution semantics~\cite{hassan2020omegalog}, (b) {\em audit logs} for OS system call events~\cite{king2003backtracking}, and (c) {\em network logs} for network traffic surveillance~\cite{datta2022alastor}.

\begin{figure*}
    \centering
    \captionsetup[subfloat]{justification=centering}
    \subfloat[Conventional (observability) \\ logging systems~\cite{ebpf,tracee,auditd}.]{
        \begin{minipage}[t]{0.22\textwidth}
            \centering
            \includegraphics[width=\textwidth]{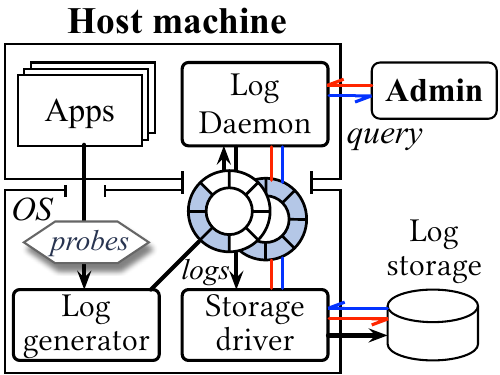}
            \vspace{-0.3cm}
            \label{subfig:overview-native}
        \end{minipage}
    }
    \subfloat[Logging systems with in-memory \\ trusted kernels~\cite{varun2023omnilog,shepherd2017emlog}.]{
        \begin{minipage}[t]{0.23\textwidth}
            \centering
            \includegraphics[width=\textwidth]{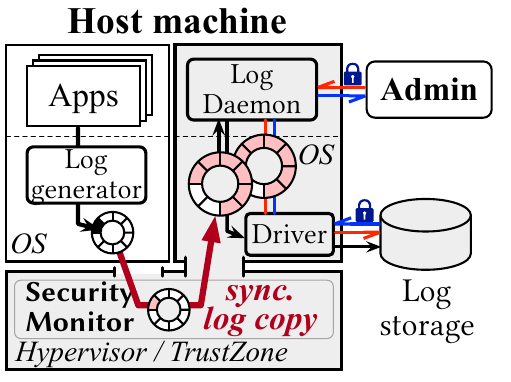}
            \vspace{-0.3cm}
            \label{subfig:overview-omnilog}
        \end{minipage}
    }
    \subfloat[Logging systems with external \\ tamper-proof audit devices~\cite{ahmad2022hardlog,sekar2024eaudit}.]{
        \begin{minipage}[t]{0.23\textwidth}
            \centering
            \includegraphics[width=\textwidth]{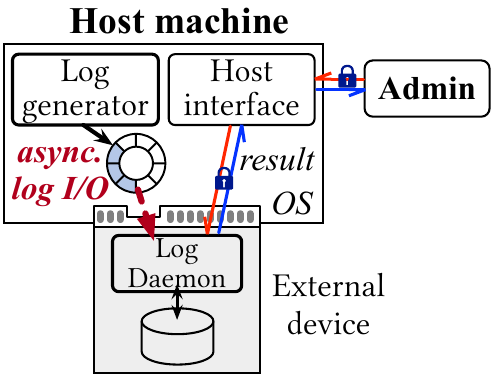}
            \vspace{-0.3cm}
            \label{subfig:overview-hardlog}
        \end{minipage}
    }
    \subfloat[\codename \\ (this work).]{
        \begin{minipage}[t]{0.23\textwidth}
            \centering
            \includegraphics[width=\textwidth]{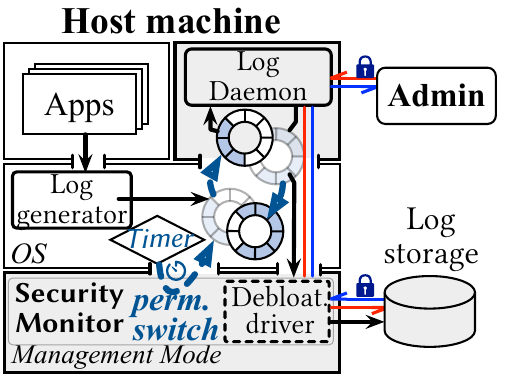}
            \vspace{-0.3cm}
            \label{subfig:overview-hhk}
        \end{minipage}
    }
    \caption{Overviews of (a) conventional observability systems and (b-d) observability protection systems (\boxhd\ : trusted components).}
    \label{fig:overview}
    \vspace{-0.2cm}
\end{figure*}

As common in both widely-deployed available software~\cite{ebpf, auditd, tracee} and the literature~\cite{pasquier2017camflow,lim2021provbpf,thomas2018camquery}, there are three main components in a conventional observability log collection stack~(\cref{subfig:overview-native}). 

First, {\em an in-kernel log generator} is to capture system events and generate logs.
Given the kernel's privileged ability to oversee and control the entire system~\cite{ebpf}, events among different layers can be captured by various {\em probes} (\eg Linux Kprobe and Tracepoint).
Second, {\em a userspace log daemon} is to manage logs.
Once captured, logs are transmitted from the kernel to the user daemon to support log management tasks (\eg filter, query, and remote retrieval).
Last, {\em a storage device} finally keeps all logs persistent locally.

\subsection{Threat Model and Assumptions}
\label{subsec:threat_model}

Like recent work in log protection~\cite{hoang2022faster,ahmad2022hardlog,varun2023omnilog,riccardo2020kennylogging,paccagnella2020custos}, we assume the enterprise host system is {\em initially benign}, but a non-privileged remote adversary starts to attack it at time $t_s$.
The adversary exploits software and kernel vulnerabilities in the host system to
(a) circumvent the common defenses~(\eg ASLR, DEP, and stack canaries) and 
(b) compromise the full system at time $t_c$.
%
Being aware of mechanisms of the system observability, the adversary attempts to tamper with the logs in kernel memory or on storage, striving to hide the attack footprints (\ie logs between $t_s$ and $t_c$)~\cite{mitrelinuxlogs}.

Before $t_c$, the observability log generator is honestly deployed inside the OS to capture logs.
Thus, all logs are initially generated and maintained correctly. For instance, the attacker cannot trick the OS into corrupting logging mechanisms in applications before $t_c$.
Post OS compromise (after $t_c$), adversaries gain full system control.
Hence, logs after $t_c$ are {\em worthless} for forensic analysis~\cite{ahmad2022hardlog,varun2023omnilog}.

{\bf \em Assumptions.}
We make standard assumptions that enterprise parties (OS vendors and administrators), computer hardware, and firmware (\eg UEFI) are trustworthy.
%
%
In addition, cryptographic protocols and key management schemes work correctly. 
Moreover, we assume the secure boot protocol is trustworthy for authenticating system configurations, ensuring the system is always benign initially.
%
We finally assume only {\em forensics-critical} logs (in \cref{subsec:system_observability}) should be protected, as they are the primary targets of the adversaries.
System display messages, such as {\em dmesg}~\cite{dmesg}, are out of scope.
Regardless, the administrator can configure the observability generator to specify what logs to be protected (\cref{sec:impl}).



\begin{table}[t]
  \caption{
  The percentage of discovered vulnerabilities in general-purpose system components and interfaces~\cite{chen2023security,cerdeira2022rezone,cerdeira2020tzsok,Shi2017DeconstructingXen} used by prior in-memory log protection systems.
  }
  \vspace{-0.2cm}
  \label{tab:general_purpose_attack_surface}
  \setlength{\tabcolsep}{3pt}
  \setlength{\aboverulesep}{-0.03pt}
  \setlength{\belowrulesep}{-0.03pt}
  \begin{adjustbox}{width=\columnwidth, center}
    
    \begin{tabular}{|cccc|ccc|}
    \hline
    \multicolumn{4}{|c|}{\textbf{Hypervisor (virtual machine manager)}} & \multicolumn{3}{c|}{\textbf{TrustZone}} \\ \hline
    \multicolumn{1}{|c|}{\multirow{2}{*}{\textbf{\begin{tabular}[c]{@{}c@{}}VM memory\\ manage.\end{tabular}}}} & \multicolumn{1}{c|}{\multirow{2}{*}{\textbf{\begin{tabular}[c]{@{}c@{}}CPU\\ virtualiz.\end{tabular}}}} & \multicolumn{1}{c|}{\multirow{2}{*}{\textbf{\begin{tabular}[c]{@{}c@{}}Code\\ emulation\end{tabular}}}} & \multirow{2}{*}{\textbf{\begin{tabular}[c]{@{}c@{}}Device\\ I/O\end{tabular}}} & \multicolumn{1}{c|}{\multirow{2}{*}{\textbf{\begin{tabular}[c]{@{}c@{}}Trusted\\ Apps.\end{tabular}}}} & \multicolumn{1}{c|}{\multirow{2}{*}{\textbf{\begin{tabular}[c]{@{}c@{}}TZ \\ kernel\end{tabular}}}} & \multirow{2}{*}{\textbf{\begin{tabular}[c]{@{}c@{}}TZ memory\\ manage.\end{tabular}}} \\
    \multicolumn{1}{|c|}{} & \multicolumn{1}{c|}{} & \multicolumn{1}{c|}{} &  & \multicolumn{1}{c|}{} & \multicolumn{1}{c|}{} &  \\ \hline
    \multicolumn{1}{|c|}{25.7\%} & \multicolumn{1}{c|}{21.5\%} & \multicolumn{1}{c|}{13.2\%} & 9\% & \multicolumn{1}{c|}{33.2\%} & \multicolumn{1}{c|}{27.8\%} & 17.1\% \\ \hline
    \end{tabular}
  \end{adjustbox}
\end{table}

\section{Motivation}
\label{sec:motivation}

\subsection{Observability Protection}
\label{subsec:existing_solutions}
An observability protection system is deployed to shield both log integrity and availability. 
%
%
%
In general, there are two main aspects to the observability log (from any sources)
protection.

%
First, a {\em log protection scheme} determines {when to shield logs} from the untrusted OS after their generation. 
This can be {\em synchronous} or {\em asynchronous}. 
In the former case, each log is secured immediately after generation (\ie blocking a syscall until its corresponding log entry is protected). 
In the latter case, logs are batched in the OS memory after generation and periodically protected.
Second, a {\em secure log environment} is required to {safely hold and manage logs} in a persistent manner.
This environment is either a combination of a secure in-memory environment and a protected disk, or only a secure disk (\eg remote storage).
%
    

\PP{Existing solutions.}
In terms of synchronous log protection, OmniLog~\cite{varun2023omnilog} is the state-of-the-art system~(\cref{subfig:overview-omnilog}). 
It blocks the system and synchronizes the log by copying it to a privileged host system layer, a so-called {\em security monitor}.
The high I/O latency makes it impractical to directly store each log entry synchronously to disk. 
%
Hence, OmniLog constructs an in-memory secure environment based on its security monitor~(\ie the hypervisor and ARM TrustZone).
Logs are kept in this environment before being eventually written to a protected disk.
%
Other solutions~\cite{dunlap2002revirt,shepherd2017emlog} follow a similar approach to building a secure environment.
Alternatively, solutions~\cite{ahmad2022hardlog,sekar2024eaudit} asynchronously protect logs by using {\em write-once-read-many} (WORM) or custom device.
%
Due to inherent I/O latencies, such systems are forced to use asynchronous protection.
%
%
In this regard, HardLog~\cite{ahmad2022hardlog}~(\cref{subfig:overview-hardlog}) is the state-of-the-art. 
It batches logs in OS memory and periodically sends them to a custom audit device, which stores logs from the host.
Such audit devices also support secure log management tasks, like remote retrieval.
Besides that, some asynchronous systems also batch and transmit logs to a network storage server~\cite{anjo2015guardat,robert2018pesos}.



\begin{table}[t]
    \centering
    \caption{Multi-layer observability log throughput, log capture overhead, and synchronous log protection~\cite{varun2023omnilog} overhead.
    \textbf{Audit Log}: system call logs~\cite{auditd};
    \textbf{App. Log}: application logs~\cite{hassan2020omegalog};
    \textbf{Net. Log}: network event logs~\cite{ujcich2021causal}.
    }
    \vspace{-0.2cm}
    \label{tab:obs_throughput}
    \setlength{\aboverulesep}{0.05pt}
    \setlength{\belowrulesep}{0.05pt}
    \setlength{\tabcolsep}{3pt}
    \renewcommand{\arraystretch}{1.0}
    \begin{adjustbox}{width=1\linewidth}
    \begin{threeparttable}
    \begin{tabular}{@{}lcccc|cc@{}}
    \toprule
    \multirow{2}{*}{\textbf{Program}}                     & \multicolumn{4}{c|}{\textbf{\#Multi-layer Observability Throughput}\tnote{\textbf{1}}} & \multicolumn{2}{c}{\textbf{Overhead (\%)}} \\ \cmidrule(l{\dimexpr0.1cm+1pt} r{\dimexpr0.1cm+1pt}){2-5}\cmidrule(l{\dimexpr0.1cm+1pt} r{\dimexpr0.1cm+1pt}){6-7} 
    \multicolumn{1}{c}{}                                  & \textbf{Audit Log}  & \textbf{App. Log}  & \textbf{Net. Log} & \textbf{Total}\tnote{\textbf{2}} & \textbf{Native-OBS}  & \textbf{\omnilog}  \\ \midrule
    \textbf{Nginx}                                        & 53,038              & 25,144                    & 594                  & 78,776         & 4.2\%             & \textbf{15.4\%}      \\
    \textbf{Redis}                                        & 40,857              & 152                       & 37,528               & 78,537         & 6.4\%             & \textbf{52.6\%}      \\
    \textbf{MySQL}                                        & 76,228              & 5,151                     & 20,583               & 101,962        & 13.8\%             & \textbf{27\%}      \\ \bottomrule
    \end{tabular}
    \begin{tablenotes}
        \item[\textbf{1}]
        {Throughput (logs/second) is counted by the conventional log generator (\textbf{Native-OBS}; \cref{sec:impl}) under the real-world workload detailed in~\cref{tab:workloads} (\cref{subsec:performance_evaluation}).}
        \vspace{0.05cm}
        \item[\textbf{2}]
        {Each log entry is around 256 bytes on average size.}
    \end{tablenotes}
    \end{threeparttable}
    \end{adjustbox}
    \vspace{-0.2cm}
\end{table}

\subsection{Limitations of Current Protection Solutions}
\label{subsec:limitations_of_existing_protection_solutions}

\PPn{L1: Risky environments with large TCB and attack surface.}
Prior work~\cite{varun2023omnilog,shepherd2017emlog,dunlap2002revirt} adopts the hypervisor or TrustZone (TZ) to protect logs in memory.
However, their bloated interfaces enlarge the vulnerabilities within the secure environment.

%
Logs are kept and managed within the trusted kernels, such as feature-rich virtual machines or the TZ secure kernels (\eg OP-TEE~\cite{optee}). 
Supporting such environments involves complex interfaces, such as CPU virtualization, code emulation, and TZ management.
Unfortunately, those full-fledged general-purpose interfaces enlarge the vulnerabilities~\cite{cerdeira2022rezone,brasser2019sanctuary,Shi2017DeconstructingXen,zhou2023coreslicing,zhang2025scrutinizer} of the secure environment.
For instance, a single vulnerability in the TZ driver interface results in the entire secure world compromise~\cite{cerdeira2020tzsok}.
%




\cref{tab:general_purpose_attack_surface} then concretely concludes the discovered vulnerabilities within general hypervisor and TZ interfaces.
%
%
Considering the severe vulnerabilities in the hypervisor, 72.07\% and 80.81\% of the common vulnerabilities and exposures (CVEs) in KVM~\cite{qumranet2007kvm} and Xen~\cite{xen} can lead to host exploitations, respectively~\cite{chen2023security}.
Meanwhile, more than 65\% of CVEs in TZ kernel-supported systems~\cite{cerdeira2020tzsok} have severe ($\ge 7$) common vulnerability scoring system (CVSS~\cite{cvss}) scores.

\PP{L2: Slow synchronous protection in log intensive scenarios.}
Modern computers can produce lots of observability logs when running realistic I/O-intensive real-world programs.
Under such scenarios, synchronously protecting every produced log, even in memory, imposes a notable slowdown.
To concretely show the slowdown, we synchronously protected logs produced by three high-performance programs: web server (Nginx), in-memory key-value store (Redis), and on-disk relational database (MySQL).
%
%
Programs are tested with default configurations and benchmark settings of {\em apachebench}~\cite{ab}, {\em memtier\_bench}~\cite{memtier}, and {\em sysbench}~\cite{sysbench}, respectively.
We employed the synchronous protection of OmniLog~\cite{varun2023omnilog}~(detailed setup can be found in~\cref{sec:performance_evaluation}).
%

%
\cref{tab:obs_throughput} illustrates the performance impact incurred by OmniLog and the conventional observability generator (Native-OBS), which stores logs within the kernel's memory, against native execution.
We observe that OmniLog's synchronous protection incurs $0.9-7.2\times$ more overhead than Native-OBS.
Due to the high throughput of observability logs, synchronous protection for each log requires frequent system interruption, involving context switches to the privileged layer (\eg changing the protection ring in x86 or the exception level in ARM), and copying the log to the protected memory inside a privileged layer, which is time-consuming.

\PP{L3: Insecure {I/O-based} asynchronous log protection window.} 
Current asynchronous protection systems~\cite{ahmad2022hardlog,sekar2024eaudit,riccardo2020kennylogging,paccagnella2020custos} expose a long time window before logs are protected to remote or local {\em tamper-proof} storage through slow I/O operations.
%
%
For instance, the state-of-the-art, HardLog~\cite{ahmad2022hardlog}, provides a lengthy \textms{15} delay.
%
%
Unfortunately, adversaries can exploit kernel vulnerabilities and obtain root access within very few syscalls~\cite{overlayfs-exp}, each of which only requires thousands of cycles, and tamper with in-kernel logs.
%



To understand the tampering problem, consider the DirtyPipe~\cite{dirtypipe} attack due to improper preservation of permissions.
Given a flaw of a simple improper \textcode{PIPE\_BUF\_FLAG\_CAN\_MERGE} flag in kernel, the vulnerability offers adversaries the capability to quickly rewrite arbitrary files (\eg \textcode{/etc/passwd}) to escalate privileges.
%
We reproduced its PoC exploit to load a kernel module to delete the logs.
We found that it took within around $12ms$ (detailed case study in~\cref{subsec:case_study}), resulting in scenarios where {\em all} attack traces were lost in \textms{15} windows.
This exploit arises from a semantic bug. Unlike common memory corruption vulnerabilities, exploiting such bugs does not require slow object manipulation or heap spraying~\cite{lin2022dirtycred}, making them brutal and ``efficient'' for adversaries.
Please note that this simple example does not show that a \textms{15} protection delay is insufficient to prevent logs from being tampered with in the majority of kernel attacks.
However, it does indicate that {PoC exploits (which are not designed to be fast) for common attacks are able to allow log tampering within a \textms{15} protection window}.
%
This underscores the necessity of reducing the protection window to substantially increase the difficulty of such attacks.

\section{\codename Overview}
\label{sec:design_overview}


%
\codename is an efficient and high-assurance observability protection system.
This section describes \codename's two key design principles (\cref{subsec:overview_secure_environment}-\cref{subsec:overview_inmemory_observability_protection}) and its system  deployment model (\cref{subsec:system_model}).
Given the rising use of ARM-based computers in both enterprise machines and cloud computing infrastructures, we use ARM as the reference architecture of this paper.
%
We further discuss the efforts of porting \codename to support other architectures in Appendix~\cref{sec:discussion}.

\subsection{\textbf{High Assurance Log Environment Design}}
\label{subsec:overview_secure_environment}
To overcome {\bf L1}, \codename protects logs in a trusted in-memory environment that is redesiged from the ground-up by leveraging three high-assurance strategies discussed in this section.

\cref{subfig:overview-hhk} shows \codename's trusted environment. 
%
%
It contains a security monitor~({\bf \codename\ Monitor or \sm}) and a protected log daemon~({\bf \codename\ Daemon or \hd}).
The monitor is required to partition and configure system resources to achieve isolation. 
It can be implemented in any higher privileged layer than the OS, \ie both the hypervisor mode (EL2 in ARM) and system management mode (EL3 in ARM).
However, \codename implements the monitor as an EL3 runtime service~\cite{uefi-runtime,atf}. 
This is because
EL3 can be kept small (and privileged) even in enterprise computers that want to run virtual machines~\cite{arm-manual}, since it would not have to include the hypervisor's TCB. 
%
%



The first strategy is to implement \sm with only {\em directly required} components for log protection, namely, {control over memory and device protection primitives}.
Memory protection is required to isolate trusted components~(\sm and \hd) from malicious CPU access by the OS. 
Device protection is required to (a) reserve a storage disk for log availability and (b) prevent malicious attacks from untrusted device access (i.e., DMA attacks).
\sm controls EL3-supported memory protection primitives for memory protection, while it leverages the System Memory Management Unit (SMMU)~\cite{varun2023omnilog,sridhara2023acai} for device protection.
The options of memory protection primitives are explained in \cref{subsec:overview_inmemory_observability_protection}, and we describe all implementation details regarding these features in \cref{sec:impl}.
%

%

The second strategy is to avoid large software components within the security monitor.
Specifically, a storage device driver is required to persist logs.
However, porting a commodity driver software into \sm is undesirable---a simple Linux SATA driver has more than 10k lines of code and complex kernel dependencies~\cite{guo2022driverlet}.
Thus, \codename implements a secure, minimized version (\cref{subsubsec:debloated_driver}).

The third strategy is to ensure log management tasks (e.g., remote retrieval) through a {\em protected daemon process} in the untrusted OS.
This avoids including trusted kernels like Linux and OP-TEE~\cite{optee} used by prior work~\cite{varun2023omnilog,shepherd2017emlog} into the system's TCB.
While a compromised OS may prevent the scheduling of the log daemon, it is unable to alter the integrity or availability of logs, nor the integrity of log management tasks such as retrieval (\cref{subsubsec:daemon_execution}). 
%


\begin{table}[t]
    \caption{Cost of different memory protection mechanisms.}
    \label{tab:prot_primitives}
    \begin{adjustbox}{width=1\linewidth}
        \setlength{\aboverulesep}{0.0pt}
        \setlength{\belowrulesep}{0.0pt}
        \begin{threeparttable}
        \begin{tabular}{cccccccc}
        \toprule
        \multirow{2}{*}{\textbf{Primitive}} & \multicolumn{7}{c}{\textbf{\#CPU cycle\tnote{\textbf{1}}\ cost per log buffer size (less means better)}}     \\ \cmidrule(lr){2-8}
        & \textbf{64B}  & \textbf{256B} & \textbf{512B} & \textbf{4KB} & \textbf{16KB} & \textbf{32KB} & \textbf{64KB} \\ \midrule
        \textbf{EL3-memcpy} & 1,785 & 2,783 & 4,680 & 24,672 & 98,978 & 198,776 & 400,903 \\
        \textbf{S2PT} & {4,482}\tnote{\textbf{2}} & {4,482}\tnote{\textbf{2}} & {4,482}\tnote{\textbf{2}} & 4,482 & 4,496 & 4,516  & 4,636       \\
        \textbf{GPT}  & {4,383}\tnote{\textbf{2}} & {4,383}\tnote{\textbf{2}} & {4,383}\tnote{\textbf{2}} & 4,383 & 4,410 & 4,398  & 4,490       \\ \bottomrule
        \end{tabular}
        \begin{tablenotes}
            \item[\textbf{1}]
            CPU runs at 1.2GHz. The cost of the Secure Monitor Call (SMC) context switch round-trip is included.
            Detailed system configuration is illustrated in~\cref{sec:impl}.
            \vspace{0.05cm}
            \item[\textbf{2}]
            S2PT and GPT's protection granularities are at the page level.
            Protecting small-sized buffers will remain at the same cost as protecting a page (4KB).
        \end{tablenotes}
        \end{threeparttable}
    \end{adjustbox}
    \vspace{-0.4cm}
\end{table}

\subsection{\textbf{Deadline-Enforced Log Permission Switch}} 
\label{subsec:overview_inmemory_observability_protection}

%
To overcome {\bf L2-L3}, \codename protects logs {\em in-memory} within short real-time deadlines set by administrators by using efficient hardware permission switches.

In-memory protection inherently provides tighter deadlines than I/O-based protection~\cite{ahmad2022hardlog}, which suffers from large device latencies and interference from other devices.
%
Naively, at configured deadlines, logs may be copied from the OS to the protected memory~\cite{varun2023omnilog}.
However, short deadlines required for security result in frequent copying.
This causes non-negligible protection overheads in scenarios where logs are frequently produced~(\eg up to 32\% runtime performance overhead as we show in \cref{subsec:performance_evaluation}).
{\em Switching} log buffer permissions using hardware permission primitives can mitigate large copy overheads.
%
%
%
To illustrate this, \cref{tab:prot_primitives} compares copying different sizes of log buffers to the privileged monitor layer (ARM's management mode or EL3~\cite{arm-manual} as used by prior work~\cite{varun2023omnilog,shepherd2017emlog}), against protecting them by (one of) hardware permission primitives\footnote{
TrustZone Address Space Controller (TZASC~\cite{tzc400}) is an alternative primitive.
However, its bus-transaction-level protection granularity lacks flexibility.
%
%
See Appendix \cref{sec:discussion} for TZASC discussion and features on other platforms.
%
%
%
} supported by \codename:

%

%
\begin{enumerate}[leftmargin=*,label=\textbullet]
    \item
    Stage-2 Page Table (S2PT), enabled by the virtualization extension, offers a second level of MMU translation which is transparent to the OS.
    It supports page-level memory access control by configuring the related table entry's permission bits~\cite{s2pt}.
    \item \vspace{0.05cm}
    Granule Protection Table (GPT) is an in-memory table structure from Real Management Extension (RME) of Confidential Computing Architecture (CCA).
    It defines page-level physical memory's accessible CPU states during MMU translation~\cite{rme-extension}.
\end{enumerate}
%


%
We find that switching permissions on log buffers (over 4KB) proves more efficient than log copying.
Copying is bound by the size of costly memory-intensive operations that are only efficient for smaller buffers (less than 512B)~\cite{du2019xpc}.
%
Hardware permission switching maintains stable costs by (a) modifying permission control registers or data structures to alter hardware permissions and (b) invalidating stale entries in the translation-lookaside buffer (TLB).
This requires significantly less CPU and memory operations.



\subsection{System Deployment Model}
\label{subsec:system_model}
We consider the same enterprise deployment model for \codename as existing research~\cite{varun2023omnilog}.
Specifically, we consider that the enterprise OS vendor (\eg Microsoft, Red Hat Enterprise Linux) will instantiate the monitor (\sm) during system boot.
%
There is a clear example of this in the past.
Microsoft Windows 11 instantiates a hypervisor-level monitor during boot-up (for kernel memory and integrity protection), and also enforces firmware-level (UEFI) extensions to support this monitor~\cite{vbs}.
We also consider that the OS vendor will modify the logging facility to invoke \sm for log protection~(\cref{subsec:design_realtime_deadline}) and management.
%
The latter include minor changes to retrieve logs remotely using the protected log daemon~(\hd) and locally from audit system tools.
Finally, for remote retrieval~(\cref{subsec:design_daemon}), we assume that the enterprise system administrator will provision cryptographic keys, as is common practice for other IT administration standards like Intel's Active Management Technology~(AMT)~\cite{intel-amt}.



\section{\codename Design}
\label{sec:design}
This section describes the main challenges towards \codename's design, and the following sections describe how we address them.

\PP{C1: Reducing software porting and deployment efforts.}
Given strict requirements to ensure high assurance, it is challenging to design an infrastructure where these strategies are satisfied without significant deployment effort (\eg completely rebuilt storage drivers for persistence or adapting secure log daemon).


\PP{C2: Enforcing short multi-layer log protection deadlines.}
Unlike prior work that only deals with OS logs, protecting logs from different layers, including applications and network logs, and under the (at least unprivileged) access of an adversary, makes it challenging to enforce uniform and very short deadlines.

\PP{C3: Delegating log management to untrusted OS securely.}
Any task naively delegated to the OS becomes untrustworthy after its compromise. 
Since it is agnostic to administrators when the OS is compromised, the OS can provide completely wrong responses after compromise even without tampering with logs.

%

\subsection{Secure Environment Software Stack}
\label{subsec:design_security_monitor}
%
%
\codename's security monitor (\sm) configures protection primitives to create isolated protection domains (\cref{subsubsec:protection_domain}).
Within the isolated protection domain, \codename synthesizes a debloated log storage device driver into \sm during offline stages (\cref{subsubsec:debloated_driver}).
Besides the debloated log driver, the monitor maintains a {\em process-based} log daemon enclave in the isolated domain by execution state and context switch protection (\cref{subsubsec:daemon_execution}).

\subsubsection{\textbf{Protection domain bootstrap and enforcement}}
\label{subsubsec:protection_domain}
\begin{figure*}[t]
    \centering
    \includegraphics[width=\linewidth]{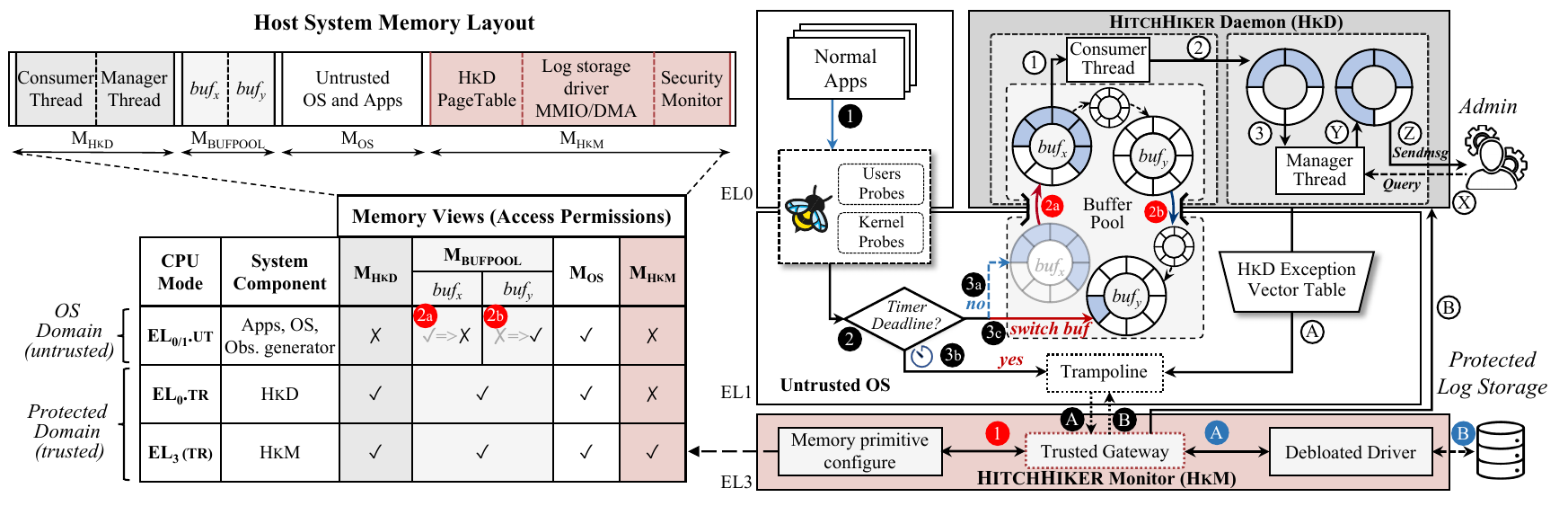}
    \caption{
    \codename memory layout and permission views (left) and workflow (right).
    The OS and applications (\boxns) are inside the OS domain (UT).
    \codename\ {\em Monitor} (\boxsm\ \sm) and {\em Daemon} (\boxhd\ \hd) are inside the protected domain (TR).
    %
    Memory views of two domains are maintained by their separated hardware primitive configurations, respectively (i.e., two S2PTs/GPTs, \cref{subsec:design_security_monitor}).
    }
    \label{fig:workflow}
    \vspace{-0.3cm}
\end{figure*}

During initialization, \sm is loaded using secure boot (\cref{sec:impl}).
%
It establishes two protection domains.
For each domain, \sm defines specific memory view permissions~(\cref{fig:workflow}-left).
\begin{enumerate}[leftmargin=*,label=\textbullet]
    \item {\em OS domain.}
    Untrusted components (i.e., OS and applications) are executing within this domain.
    They can only access their own memory, as well as shared memory and unprotected log buffers (which are dynamically protected, explained in next paragraphs).
    
    \item {\em Protected domain.}
    Trusted components (i.e., \hd and \sm) are executing within this domain.
    \hd contains two threads~(whose role is explained in~\autoref{subsec:design_realtime_deadline}) and it is allowed to access all log buffers.
    \sm manages its own code and data, as well as the device driver interface (\cref{subsubsec:debloated_driver}) and \hd's protected execution states and page table (\cref{subsubsec:daemon_execution}) within the protected domain as well.
    
\end{enumerate}

%


%
%

%

%
%
Depending on the hardware protection primitive, each domain's memory view is enforced by {\em maintaining separate S2PTs or GPTs}.
In particular, GPT configuration is as follows.
In the OS domain's GPT, only untrusted components are marked as {\em normal-world accessible}~\cite{rme-extension}.
\hd memory region is marked as {\em none-accessible}, while \sm is marked as {\em root-world-assessible} (EL3-exclusive).
When switching to the protected domain's GPT, \hd memory is marked as {\em normal-world accessible}.
Such a design is agnostic to the Realm world~\cite{rme-extension-website} and, therefore, does not interfere with GPT's original functionality (i.e., supporting confidential VMs in Realm world~\cite{cca}).
Configurations of S2PT are similar to GPTs---the OS domain's S2PT only grants access to the untrusted components, while the other can access trusted components~\cite{hof2022blackbox}.

With enforced memory views, \sm dynamically protects log buffers by removing the OS domain GPT or S2PT's access permission to buffers.
CPU cores that execute in the OS domain share GPT (or S2PT) entries in TLB~\cite{rme-extension}.
As such, a synchronized memory view is maintained for the OS domain, even though different cores may concurrently protect log buffers (e.g., by requesting \sm to change the OS domain's GPT entries) at runtime (\cref{subsec:design_realtime_deadline}).

To switch between domains (e.g., to execute \hd or protect log buffers), the OS executes a Secure Monitor Call (SMC~\cite{smccc}) to \sm.
%
%
At such calls, \sm validates the request and switches the GPT/S2PT to perform the requested functionality (e.g., log protection).
A detailed list of SMC interfaces implemented by \codename and their validations can be found in \cref{tab:smc} (Appendix \cref{appen:vul_mitigation}).

Note that, in addition to GPT/S2PT, \sm leverages System Memory Management Unit (SMMU) to prevent DMA attacks against the protected domain. 
The SMMU limits the addressable physical memory ranges of a device~\cite{sridhara2023acai}. 
SMMU configuration interfaces (memory-mapped I/O interfaces and stream tables) are only accessible within the protected domain (by aforementioned memory permission restrictions).

%
%

%

\vspace{-0.1cm}
\subsubsection{\textbf{Debloated driver synthesis and protection.}}
\label{subsubsec:debloated_driver}
\codename synthesizes a debloated driver prior 
to system bootstrap, which is later instantiated into \sm during initialization.
%
%
Driver debloating is achieved by record-and-replay mechanisms~\cite{wang2022rttee,guo2022driverlet} for storage drives (\eg SATA, USB, etc), as a one-time effort.

Considering a block storage device as a finite state machine, driver operations are distilled into a sequence of {\em driver-device} interactions, each of which is a state transition.
These interactions remain consistent irrespective of variations in I/O workloads~\cite{guo2022driverlet}.
%
During recording, sample I/O jobs are issued to log the storage to templatize the following interactions (detailed in \cref{sec:impl}):
(1) the sequences and content of Memory-mapped I/O (MMIO) write operations to the log storage, 
(2) the allocation of direct-memory-access (DMA) buffers for the log storage, and 
(3) interrupt request (IRQ) numbers invoked by the log storage, along with the completion status.
%
The template is then installed into \sm.
%


%
\sm sets up and protects the driver from the template during initialization to enable communication with the log storage disk.
In particular, the driver's MMIO regions (which are fixed addresses) are only mapped to the {protected domain}.
Similarly, DMA buffers to hold I/O payloads are pre-allocated in the protected domain.

%
%
Thus, the untrusted OS cannot manipulate the driver interface.
To ensure interrupts from the log storage disk are routed to the protected driver only, \sm updates the IRQ table in the EL3 interrupt management framework~\cite{irqroute}.
At runtime, whenever a log persistence or deletion command is issued, \sm replays the template with dynamic parameters (\eg MMIO control sequences with payloads filled in the DMA buffers), to directly read or write the corresponding log disk sectors.

\vspace{-0.1cm}
\subsubsection{\textbf{Protected userspace daemon execution}}
\label{subsubsec:daemon_execution}
\codename enables loading an unmodified program into a protected userspace process (or enclave) as \hd.
%
This allows administrators to build log management functionality easily.
This section describes how \hd's memory and execution state are protected in the protected domain with secure context switches.
%
%
\cref{subsec:design_daemon} explains how log management tasks are achieved using \hd.
%

\PP{Memory and execution state protection}
%
%
%
\hd (including its memory and execution stack) is loaded into the protected domain, in a manner inspired by SGX ~\cite{mckeen2013sgx} and recent research~\cite{hof2022blackbox,hofmann2013inktag,zhang2023shelter}.
%
%
%
%
Specifically, during machine provisioning, the binary's SHA-256 integrity hash is pre-installed onto protected storage.
During system boot, the OS allocates its process context, loads the binary within a reserved region, and requests \sm to attest to its initial integrity (\ie via the hash).
To prevent {\em time-of-check-to-time-of-use} attacks, the monitor protects the process region (by the aforementioned protection domain) before attestation.
%
%

%
After attestation, \hd's pages are {\em pinned} to memory (\ie no swapping). 
The monitor then copies the page tables of the process into its own memory region.
As \hd's page table is maintained inside the monitor's memory, the OS cannot make any changes to verified mappings, thereby preventing page remapping attacks~\cite{hofmann2013inktag}.
Pinning is acceptable since \hd would typically require a small amount of memory~(less than 8MB in our implementation).
%

%
\hd is initialized with two threads: a consumer thread and a manager thread.
The former thread consumes protected logs, while the latter issues log persistence operations (\cref{subsec:design_realtime_deadline}) and securely interacts with remote tasks (\cref{subsec:design_daemon}).
%


\PP{Secure context switch handling}
Like any user process, the task scheduling of \hd is controlled by the OS (untrusted).
However, the OS cannot directly transition into \hd's context, as it is in the protected domain.
Hence, whenever the OS wants to schedule these threads, it calls the monitor using SMC (\cref{subsubsec:protection_domain}).
In such calls, \sm restores \hd's context state and memory permission view.

Once scheduled, \hd cannot exit directly to the OS for system calls or interrupts due to protection domain enforcement.
To {\em securely and transparently} support \hd exits, \sm enables a custom exception vector table when executing \hd, in which all exceptions are interposed by a trampoline SMC (\cref{fig:workflow}, \circlewhite{A}--\circleblack{A}).
Thus, at \hd exits, the monitor saves \hd's context state before exiting to the OS (\circleblack{B}).
Whenever the OS calls the monitor to reschedule \hd, the context is restored (\circlewhite{B}).
\sm ensures that system call results are sanitized to prevent IAGO attacks~\cite{stephen2013iago} (\eg OS returns a malicious pointer).
We explain how network system calls are sanitized in~\cref{subsec:design_daemon}.
In the future, if other system calls, such as file system support, are needed by the program, a TCB-conscious library operating system (LibOS)~\cite{tsai2017graphene,shinde2017panoply} can be integrated with \hd. 
%

\subsection{Deadline-Enforced Unified Log Protection}
\label{subsec:design_realtime_deadline}

%
\codename\ {\em enforces} a short log protection deadline by employing a periodic hardware timer by the OS.
The administrator configures a real-time protection delay of $T_p + \epsilon$, where $T_p$ is the timer interrupt frequency and $\epsilon$ is the latency incurred while protecting logs in memory.
Modern hardware timers have a high resolution~\cite{wang2022rttee}. 
Therefore, $T_p$ can be tiny (\eg \textus{100}--\textms{1}).


%
This section first explains how different logs are collected in a unified location to standardize the protection window.
Then, it explains how the timer ensures logs are protected in memory within a set deadline.
Finally, it describes how protected logs are securely persisted in the background to guarantee eventual availability.
%

%

%

%


\subsubsection{\textbf{Unified log collection in kernel buffer pools}}
\label{subsubsec:design_log_collection}
Capturing logs at different locations (\eg network drivers, applications) in an ad-hoc manner to protect them can increase protection delay.
\codename avoids this by unified capturing logs within the kernel before protection~(\circleblack{1}--\circleblack{2} in~\cref{fig:workflow}).
It requires two aspects for unification: (a) a central region to hold all logs and (b) a mechanism to generate the captured events as logs to that region.
To create a central region, \codename allocates a {\em pool of per-core} buffers within the kernel observability generator's memory.
For each kernel thread, logs are always written to the {\em current} per-core buffer within the pool, thereby avoiding cross-core contention as multiple kernel threads can write logs simultaneously.
A per-core pool of (at least two) buffers is required, because \codename dynamically switches the {\em current} buffer's permission under the OS's domain for protection (explained in \cref{subsubsec:design_log_protection}).
Hence, there should always be an available buffer as the {\em current} to the OS.
%
%
The size of the buffers is administrator-configured, and the buffers are pre-allocated during system initialization.
If the buffer becomes full, very rare in honest scenarios based on proper configuration in our evaluation~(\cref{sec:performance_evaluation}) and prior work~\cite{varun2023omnilog}, the generator waits for a buffer to be consumed and returned.
This prevents attackers from overwriting their attack trace even if they generate many spurious logs from userspace~\cite{riccardo2020kennylogging} before full system compromise.
%


\codename captures the logs (\circleblack{1}) by extended Berkeley Packet Filter~(eBPF~\cite{ebpf}), a widely-used generator~\cite{agman2021bpfroid, minna2023sok} in prior work (see implementation details in \cref{sec:impl}).
A challenge in capturing {\em application logs} is locating and automatically intercepting the routines where logs are generated.
\codename addresses this by using contextual tracking of file system calls.
In particular, applications leverage the OS file system to write logs. 
Thus, the generator records {\em write} syscalls and extracts file paths in kernel structures (\eg \textcode{dentry} in \textcode{vfsmount}) to match them with administrator-defined log file paths (\eg \textcode{/var/log})~\cite{tracee}. 
For log file writes, \codename extracts the log string using an eBPF helper \textcode{bpf\_probe\_read\_user}.
%
%
For logs from network and system calls, \codename injects BPF programs into kernel tracepoints to attach all 436 system calls and socket-related events.
\codename follows Linux Auditd's format~\cite{auditd} to generate system call log entries, which include parameters, return values, and other fields (\eg process identifiers and timestamps).
For network traffic logging, \codename logs IP, ports, and other packet data from the kernel socket structure of network traffic.
%

%






\vspace{-0.1cm}
\subsubsection{\textbf{Controlled hardware timer-based buffer protection.}}
\label{subsubsec:design_log_protection}
The OS deploys a hardware timer based on configured frequency $T_p$ to direct a kernel thread to initiate log protection.
As the OS is initially benign, it will correctly deploy the timer before compromise.
%
%
To ensure the short protection window, \codename~{\em controls} the delay after the timer expires (each $T_p$) to ensure that protection is achieved in an isolated small period ($\epsilon$).
All these aspects are explained in the following paragraphs. 
Once logs are captured, \codename writes them to the {\em current} buffer~(\cref{fig:workflow}, \circleblack{\Aa{3}{a}}). 
When a timer deadline is reached, a kernel thread calls \sm (\circleblack{\Aa{3}{b}}--\circleblack{A}) to temporarily switch the OS access permission to the current buffer, marking the buffer as inaccessible to the OS domain (\circlered{1}--\circlered{\Aa{2}{a}}).
After protection is complete and execution returns to the OS, a new buffer is fetched from the pool to be the {\em current} buffer (\circleblack{B}--\circleblack{\Aa{3}{c}}).
The previous (now protected) buffer will be re-inserted into the pool after consumption~(\cref{subsubsec:design_log_persistence}).


To correctly capture logs and protect them by constant timer preemptions, both log buffer writing and protection are executed in {\em atomic contexts}.
Otherwise, the preemptions between log writings and protections will not only disrupt the deadline but cause faults (\eg by writing logs to a just-protected buffer before fetching a new current buffer).
To this end, \codename enforces each core to acquire a lock before every log buffer write.
On timer interruption, the timer thread acquires those locks before protection. 
Since the lock is released after every log buffer write, this timer thread acquires it within a tiny delay (less than a hundred CPU cycles on average in our experiments).

\codename forcibly isolates $\epsilon$ (the time delay between the timer thread invocation and log protection finished by \sm) to a small value.
%
Otherwise, adversaries may significantly extend $\epsilon$ to corrupt the protection deadline.
This property is regulated by (a) prioritizing the timer interrupt kernel thread and (b) enforcing the aforementioned atomic execution of the log protection procedure.
%
The former (interrupt prioritization) prevents the timer thread invocation from being delayed by other interrupts (\eg by raising spurious interrupts~\cite{lee2021exprace}). 
%
%
The latter (atomic permission switching execution) ensures the permission switch procedure execution cannot be extended by other task preemptions or IRQs.

%
\subsubsection{\textbf{Fast buffer consumption with eventual persistence.}}
\label{subsubsec:design_log_persistence}
\hd consumes the protected buffer (\ie persists its contents to disk) and then returns the buffer into the kernel buffer pool (by permission switching).
It consumes buffers efficiently using multi-threading and {\em real-time} techniques~\cite{ahmad2022hardlog}.
Since \hd might not be scheduled after OS compromise, \sm ensures all remnant buffers (in the protected domain) are persisted before shutdown events.

\hd separates its threads~(\cref{subsubsec:daemon_execution}) to the consumer thread for log consumption and the manager thread for log persistence and admin task response.
By doing so, heavy I/O and network jobs are dispatched to the manager thread.
In addition, the consumer thread is prioritized by (an honest) OS over other threads.
Moreover, all hardware interrupts, except for the protection scheme timer, are disabled by the OS during the thread's execution.
\hd requests the monitor to persist the consumed logs through the debloated driver.
%
Since \hd is at userspace, it cannot directly issue I/O operations.
%
%
Hence, the manager thread executes an {\em ioctl} system call as a proxy, which then invokes \smc to request \sm.
This is just a notification mechanism, all message passing between \sm and \hd is done using a reserved {\em communication buffer} within \hd's memory.
When there are no logs, \hd goes to sleep.
The monitor intercepts power management operations to ensure all protected logs are persisted.
%
%
In particular, it intercepts the Power State Coordination Interface (PSCI) and issues an I/O command to persist all protected logs forcibly before power events~\cite{varun2023omnilog,ahmad2022hardlog}.
%

\cref{fig:workflow} illustrates the procedure.
The consumer thread consumes logs in the protected buffer (\circlewhite{1}--\circlewhite{2}), and asks the monitor (\circlewhite{A}--\circleblack{A}) to edit the OS domain's permission to return the buffer (\circlered{1}--\circlered{\Aa{2}{b}}).
Meanwhile, the manager thread requests \sm to persist them into secure storage via the debloated driver interface (\circleblue{A}--\circleblue{B}).
Finally, before power events, \sm issues an I/O command via its debloated driver to force all protected logs to be persisted to the storage.




\subsection{Secure Delegated Remote Log Management}
\label{subsec:design_daemon}

\codename empowers \hd to securely support remote log management tasks, such as retrieval, prior to OS compromise.
After compromise, the integrity of any completed management task is ensured, and the OS can only prevent new task completion~(\eg not schedule \hd).
To achieve such guarantees, \hd sets up a secure channel to a remote administrator, using provisioned cryptographic secrets, over a sanitized OS-delegated network transport.
%

\subsubsection{\textbf{Daemon secret establishment.}}
%
%
\label{subsubsec:design_daemon_secret}
At machine provisioning~(\cref{subsec:system_model}), the IT administrator installs the log daemon \hd, and sets up cryptographic secrets for runtime secure communication.
%

%
%
To ensure secure communication, the administrator provisions a key pair.
The secret portion of the key is installed on the protected disk, while the public portion is kept by administrators for later authentications.
%
%
Besides that, admins also install their own public key into the protected disk to ensure a {\em two-way} authenticated and secure channel can be created.
Since the protected storage disk can only be accessed by the monitor,
during \hd loading, \sm securely installs the provisioned keys within a reserved region of \hd.
Thus, the key remains protected from the untrusted OS.
%

%


%

\subsubsection{\textbf{Secure channel on sanitized delegated network.}}
\label{subsubsec:design_daemon_syscall}
\codename sends and receives packets through OS network interfaces, but ensures packets are secured end-to-end between \hd and the administrator.
This is ensured by the Transport Layer Security (TLS) protocol and the key pair provisioned.
Therefore, adversaries cannot impersonate the remote admin (\eg forge and issue log deletion requests and responses).
The remaining paragraphs illustrate how \hd network system calls are supported and sanitized.

\sm interposes \hd's network service system call invocations and returns (\ie context switches) to perform secure domain interpositions.
The interposition is enforced through \hd's specialized exception vector table, which is transparent to the OS (\cref{subsubsec:daemon_execution}).
As such, \hd's execution context is securely saved and managed by \sm.
At syscall invocations, before restricting the memory permission view to enter the OS, syscall pointer parameters are redirected to a shared buffer that is accessible in the OS.
Otherwise, the OS (within in restricted permission view) cannot access \hd's memory buffer pointed by the syscall parameters.

Once the syscall returns, the monitor sanitizes the return values before resuming \hd.
In particular, to prevent pointer-based IAGO attacks~\cite{stephen2013iago}, \sm ensures that no pointer is returned for a region that belongs to \hd's domain.
Moreover, \sm copies the packet data from the shared buffer into \hd's buffer.
Then, \sm transits to the protected domain (enables \hd's memory permission view), restores the context, and enforces its maintained \hd page and vector tables.
Finally, \sm transits the CPU execution to \hd.

\cref{fig:workflow} (\circlewhite{X}--\circlewhite{Z}, \circlewhite{A}--\circlewhite{B}, \circleblack{A}--\circleblack{B}) illustrates the interposed context switch.
Whenever \hd requests network services (\circlewhite{X}, \circlewhite{Z}), it traps to \sm (\circlewhite{A}--\circleblack{A}) by its customized exception vector table.
\sm then saves its context, redirects pointer parameters, transits to the OS domain (\circlered{1}), and then exits to the OS (\circleblack{B}).
Whenever the OS finishes the service, \sm interposes the return (\circleblack{A}) and performs the context switch on the OS' behalf.
It restores the saved contexts, enables \hd's domain (\circlered{1}), enforces its page and vector table, and directly returns to \hd's execution context (\circlewhite{B}).


\section{Implementation}
\label{sec:impl}
%

%
We prototyped \codename for an ARM-Juno R2 board featuring six CPU cores~(a dual-core Cortex-A72 and a quad-core Cortex-A53) alongside 8GB SDRAM.
A 256GB SATA WD BLUE SSD was connected to the board for secure log storage, and a 1TB Sandisk Extreme Pro SSD was used by the OS for normal filesystem storage.
The board runs Linux 5.3 as the OS, using Trusted-Firmware-A (TF-A) arm\_cca\_v0.3~\cite{atf} at EL3 as the base of \sm.
%
%
%
For the GPT implementation, there is no commercial hardware that supports Realm Management Extension (RME).
To validate the functionalities of \codename, we developed a functional prototype on ARM Fixed Virtual Platform (FVP) with RME support.
To further emulate the performance of GPT on our board, we followed common practices (described in the following paragraphs).
%
\cref{tab:impl_breakdown} offers a breakdown of source code lines modified or introduced.

\begin{table}[t]
    \caption{\codename component implementation breaks down in source lines of code (LoC).}
    \label{tab:impl_breakdown}
    \setlength{\aboverulesep}{-0.05pt}
    \setlength{\belowrulesep}{-0.05pt}
    \setlength{\tabcolsep}{3pt}
    \begin{adjustbox}{width=\columnwidth, center}
    \begin{tabular}{@{}lrr@{}}
    \toprule
    \textbf{Component}           & \textbf{Base}  & \textbf{SLoC} \\ \midrule
    \textbf{\codename Monitor (\sm)} &                & \textbf{2.7K in total}   \\
    \ \ \ \ Hardware permission - GPT configuration    & TF-A~\cite{atf}           & 399           \\
    \ \ \ \ Hardware permission - S2PT configuration    & TF-A~\cite{atf}           & 738           \\
    \ \ \ \ Debloated driver             & Sata\_sil24~\cite{sata-sil24}    & 566           \\
    \ \ \ \ \hd context interposition            & TF-A~\cite{atf}           & 1,106         \\
    \midrule
    \textbf{Operating System}   &                & \textbf{3K in total} \\
    \ \ \ \ Observability generator      & Tracee~\cite{tracee}, Libbpf~\cite{libbpf} & 860           \\
    \ \ \ \ Timer and buffer pool management      & Linux kernel module  & 1,618   \\
    \ \ \ \ \hd binary loader       & Linux kernel module  & 584   \\
    \midrule
    \textbf{\codename Daemon (\hd)}   &                & \textbf{0.8K in total} \\
    \ \ \ \ Consumer thread         & --     &  208           \\
    \ \ \ \ Manager thread          & --      & 622              \\ \bottomrule
    \end{tabular}
    \end{adjustbox}
\end{table}

\PPn{\sm and protection domain enforcement.}
\sm's bootloader images are burnt into the {\em secure boot region} (trusted boot ROM and SRAM), which is not accessible to the normal OS once boot.
During boot, the host memory is partitioned, specified by TF-A (\textcode{bl\_regions}) and device tree.
Contiguous physical memory regions are reserved for \hd and the log buffer pool by specifying the Linux contiguous memory allocator~\cite{li2021twinvisor}.
All the partitions (their ranges) can be easily adjusted and configured by administrators.

We reserve memory in TF-A to maintain two GPT/S2PT table instances.
%
%
For S2PT implementation, we set up \textcode{HCR\_EL2.VM} during boot to enable the second-level address translation.
All S2PT control interfaces (\textcode{VTCR\_}, \textcode{VTTBR\_EL2}) are directly configured by TF-A.

We emulate GPT's performance by following common practice~\cite{zhang2023shelter,sridhara2023acai,li2022ccakvm}.
%
%
First, we emulate the GPT control interface cost by substituting all control registers (\textcode{GPCCR\_}, \textcode{GPTBR\_EL3}) with idle EL3 ones.
Second, we program GPT table structures per GPT specifications~\cite{rme-extension} by using TF-A official code.
Third, we flush the entire TLB entries after every GPT configuration change.
Even if the cost of hardware GPT checks (expected to exhibit good caching behavior) is not included, this {\em will not affect the relative overhead} since GPT checks are applied to the full system~\cite{li2022ccakvm}.

\PPn{Debloated driver synthesis.}
To synthesize a debloated driver offline, we used the default SATA driver (\textcode{sata\_sil24})~\cite{sata-sil24} on our system for recording.
%
%
In principle, any driver can be used.
The recording is performed by sample I/O that writes directly to the log device (\eg \textcode{/dev/sda}), with flags \textcode{O\_DIRECT} | \textcode{O\_SYNC} to bypass the file system (\eg journaling) and block abstractions.
To record MMIO and DMA operations, our recorder instruments the lowest kernel functions (\textcode{readl}, \textcode{writel}, and \textcode{dma\_alloc}) before I/O.
%
%
During debloated driver setup in \sm, its device IRQ is isolated from the OS (\textcode{plat\_ic\_set\_interrupt\_type}) and routed to \sm by setting the \textcode{SCR\_EL3.FIQ} bit in TF-A (\cref{subsubsec:debloated_driver}).

\PPn{\hd and timer interrupt prioritization.}
\hd's initialization is supported by a customized binary loader (in a kernel module).
Its thread execution stacks and specialized exception vector table are reserved during \sm setup (\cref{subsubsec:daemon_execution}).
%
%
To prioritize the protection timer interrupt over others, we utilized TF-A's interrupt management framework~(the function \textcode{plat\_ic\_set\_interrupt\_priority}).
In particular, we programmed the Interrupt Priority GIC MMIO registers (\textcode{GICD\_IPRIORITYR}) to grant the timer interrupt with the highest priority~(GIC\_HIGHEST\_NS\_PRIORITY) in the normal world.
%
%
Note that the kernel (at EL1) can also access this register. 
Currently, we do not protect attackers from tricking the vulnerable kernel into changing the timer configuration. 
We discuss such attacks and techniques to enable full timer protection in \cref{sec:discussion}.



%
%

\PP{Observability generator.} 
The generator is built and extended upon Aqua Tracee~\cite{tracee}, a widely deployed eBPF-based forensics tool~\cite{agman2021bpfroid, minna2023sok}.
We extended the Linux BPF helpers with a new one named \textcode{bpf\_write\_current\_buf} and integrated it into libbpf.
The helper's underlying kernel function unifies log generation into our {\em current} kernel buffers (\cref{subsubsec:design_log_collection}).
%
%
Deploying and verifying such an extended BPF helper is straightforward and aligns with common practices~\cite{lim2021provbpf}, requiring \textasciitilde 50 lines of change for the BPF verifier. 

\section{Security Evaluation}
\label{sec:sec_evaluation}




\subsection{Security Claims and Analysis}
\label{subsec:security_analysis}

We analyze \codename's log integrity and availability guarantees through a sequence of security claims ({\bf S1}--{\bf S5}).

\PP{S1}
{\em
All observability logs before OS compromise are captured.
}

Prior to system compromise, the log generator operating within the OS faithfully captures all (OS, application, and network) logs within the {\em current} kernel log buffer.
The administrator configures the buffer pool to allocate enough space (\cref{sec:performance_evaluation}) for logging.
If buffers are exhausted within the pool, the generator will pause the event and wait for a buffer to be consumed by \hd~(\cref{subsubsec:design_log_collection}).
%
Hence, the attacker cannot prevent events from being logged, and neither can they flood the buffers with numerous spurious event logs to force the system into dropping previously-logged entries~\cite{riccardo2020kennylogging}.
%


%



\PP{S2}
{\em 
All collected logs will be protected in memory within the configured, short real-time deadline ($T_p + \epsilon$).
}

The OS, which is initially benign, faithfully configures a hardware timer based on the frequency ($T_p$) specified by the administrator.
This cannot be modified by the attacker until they compromise the OS.
Once a timer interrupt is fired, a kernel thread calls \sm (using an SMC) to protect logs.
The delay between when the interrupt is fired and logs are protected ($\epsilon$) is kept small and regulated because (a) the buffer permission switch procedure being executed in an atomic context with preemption and IRQ disabled, (b) the timer interrupt handling is always prioritized (\cref{subsubsec:design_log_protection}), and (c) buffer permission switch is fast.
As such, adversaries cannot prolong either $T_p$ or $\epsilon$ via launching malicious interrupts~\cite{lee2021exprace} or flooding the buffer with numerate events~\cite{riccardo2020kennylogging} (further validated in~\cref{subsec:protection_scheme_evaluation}).
%


%
%

%

\PPn{S3.}
{\em
All protected logs in memory cannot be accessed by the OS.
}

Once the log protection \smc is handled by \sm (\ie at $T_p + \epsilon$), the log buffer's memory access permission under the OS domain is removed.
The log buffer is also protected by SMMU; hence, device-based DMA attacks are prevented.
\sm uses the same mechanisms---memory permission and SMMU---to protect itself and the data structures of SMMU from the OS~~(\cref{sec:impl}).
Hence, any malicious access to the protected domain is prevented.
%



%
\PP{S4}
{\em
All protected logs in memory will be persisted in the secure log storage disk before system shutdown.
}

\hd's consumer thread, which is securely loaded~(\cref{subsubsec:daemon_execution}), asynchronously issues I/O commands through \sm's debloated driver interface to persist protected logs (\cref{subsubsec:design_log_persistence}).
Prior to compromise, the OS will schedule the consumer thread to ensure logs are persisted.
%
After compromise, the OS cannot modify \hd's operations due to domain enforcement, and it can only refuse to schedule the thread.
If this happens, the logs will remain in protected memory until a system shutdown event is triggered by the OS.
%
Assuming no power or hardware failures (which cannot be controlled by our attacker), \sm will intercept all shutdown events and leverage its protected driver to persist all remnant in-memory logs to disk.
%



%
\PP{S5}
%
{\em
All stored logs will only be deleted (e.g., after retrieval to remote storage) by the system administrator.
}

A compromised OS cannot directly send log deletion commands to the protected disk, since the disk is isolated, and can only be accessed by the debloated driver in \sm~(\cref{subsubsec:debloated_driver}).
\sm only accepts log deletion requests from \hd, which will only send such a request if it receives a command from the system administrator.
In particular, \hd and the system administrator leverage a cryptographic key pair---securely provisioned in the protected disk~(\cref{subsubsec:design_daemon_secret}) and loaded by \sm during initialization~(\cref{subsubsec:daemon_execution})---to establish a secure, authenticated secure channel through the OS-controlled network (\cref{subsubsec:design_daemon_syscall}).
Since the OS cannot extract or modify these keys, it cannot compromise the channel (\eg impersonate the administrator).
Even though the request notification between \hd and \sm is relayed by the OS (using an SMC), the request parameters (\eg which disk sectors to clear) are protected in \hd's memory~(\cref{subsubsec:design_log_persistence}).
\subsection{Protection Deadline Analysis}
\label{subsec:protection_scheme_evaluation}

This section describes the {worst-case} deadlines ($T_p + \epsilon$) found for in-memory log protection, given different configurations of timer frequency ($T_p$) on our test system.
%


%
\PP{Settings.}
%
%
We ran two workloads.
The first workload is from a stressful benchmark, {\em getpid-flood}, based on prior work~\cite{ahmad2022hardlog}. 
This benchmark executes one million \textcode{getpid} system calls in a loop to intensively generate logs.
%
%
The second workload is from a real-world program, {\em Nginx}, which produces diverse logs.
For Nginx, we used benchmark \textcode{ab}~\cite{ab} under settings in \cref{tab:workloads}.
%
%
We evaluated \codename in three configurations of $T_p$: \textms{1}, \textus{500}, and \textus{100}.
Both S2PT and GPT are employed as the \codename's permission primitive, respectively, and we report the worst-case experimental results. 
The experiment was repeated 100 times.
%

\begin{figure}[t]
  \centering
  \includegraphics[width=0.9\linewidth]{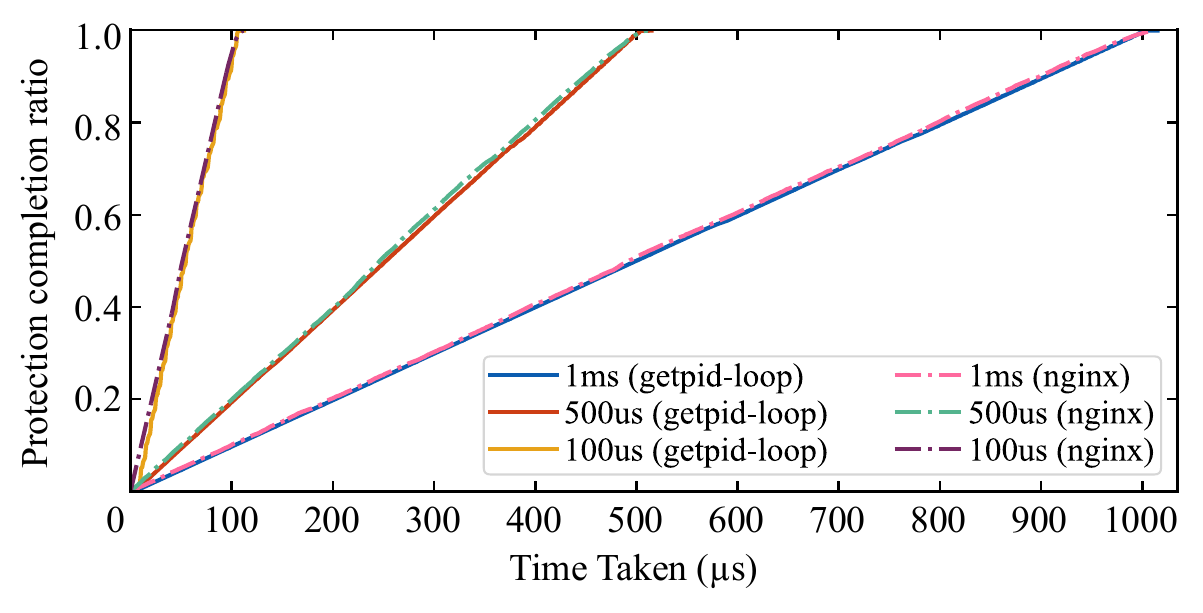}
  \vspace{-1em}
  \caption{
CDF shows the relation between different protection timer $T_p$ (\textms{1}, \textus{500}, and \textus{100}) deadline settings and \codename's actual in-memory protection window (delay).}
  \label{fig:policy_cdf}
  \vspace{-1em}
\end{figure}

\noindent \textbf{Results.}
\cref{fig:policy_cdf} shows the cumulative distribution function (CDF) of the time taken to protect logs with different $T_p$.
Under the {\em getpid-flood} benchmark, when $T_p$ is assigned values of \textms{1}, \textus{500}, and \textus{100}, 99.6\%, 99.1\%, and 92.8\% of logs, respectively, were protected in a window less than the assigned $T_p$.
Each configuration had an additional delay ($\epsilon$) in the worst case, which ranged between \textus{12.12} and \textus{14.96}.
%
%
In the case of {\em nginx}, 99.5\%, 99.3\%, and 95.5\% of logs were protected within $T_p$ (ranging from \textms{1} to \textus{100}, respectively).
The worst-case values of $\epsilon$ ranged between \textus{11.52} and \textus{13.73}.
As Nginx's workloads also generate considerable logs (around 79k per second), its CDF is similar to getpid-flood (which generates logs at around 178k per second) but slightly better.
Under both settings, the delay $\epsilon$ was negligible and stable, due to \codename's enforcement mechanisms~(\cref{subsubsec:design_log_protection}) and fast permission switches (\cref{subsec:overview_inmemory_observability_protection}).

%




%
Recall that the state-of-the-art asynchronous protection system~\cite{ahmad2022hardlog} incurs a best-case deadline of \textms{15}.
Hence, with configurable deadlines between roughly \textms{1.012} and \textus{115}, \codename exposes $93.3\%-99.3$\% shorter attack windows.
The next section shows how such short protection windows further prevent log tampering attacks when compared to prior work.
%






\subsection{Log Tampering Case Study}
\label{subsec:case_study}
%
%
To evaluate whether \codename's short deadlines provide a significant hurdle for adversaries to tamper with logs, we simulated a \textit{powerful local attacker} to delete logs in memory as fast as possible.

\PP{Attack setup.}
We consider powerful local attackers who know the log buffers' physical addresses.
Attackers also {\em already held} the foothold to access the system (at time $t_s$).
Such settings avoid time to establish footholds.
To begin with, attackers exploit kernel vulnerabilities to escalate privileges to root at $t_c$.
Upon system corruption, attackers halt the log generator and delete logs in memory.

%
We reproduced ten kernel CVEs with PoC exploits to escalate privileges to root.
%
CVEs were selected from well-known security assessment forums (\eg\ {\em Seclist, Openwall}) with high Common Vulnerability Scoring System (CVSS)~\cite{cvss} scores.
Also, they were collected from two categories~\cite{chen2011linuxkernel}: (i) semantic bugs like improper permissions or security checks, and (ii) memory corruptions (\eg stack or heap-related temporal and spatial corruptions).
%

%
%

To simulate a powerful attacker, we chose the fastest out of the ten exploits on our host machine (\cref{sec:impl}): DirtyPipe~\cite{dirtypipe}.
The full list of reproduced CVEs and their execution times are provided in Appendix~\cref{appen:all_attacks}.
After gaining root, the attacker launches a malicious kernel module (\textcode{.ko}) to \textit{clear logs residing in the unprotected buffer}, erasing the attack evidence.
During the attack, \codename was fully deployed with different protection timer configurations $T_p$ of \textms{15}, \textus{1}, and \textus{100}. 
We ran this attack 50 times by activating the attack script at random times (same as the~\cref{subsec:limitations_of_existing_protection_solutions}).

\begin{table}[t]
  \caption{
  Local attack time and log event statistics. 
  \#Total Logs denotes the number of all log events captured during the attack.
  \#Lost Logs shows the number of events that are cleared by the attacker (under different $T_p$). 
  The account under $T_p$ \textms{15} denotes the number of tampered logs under state-of-the-art asynchronous protection solution, HardLog~\cite{ahmad2022hardlog}.
  }
  \label{tab:demo_attack}
  \setlength{\tabcolsep}{3pt}
  \setlength{\aboverulesep}{-0.03pt}
  \setlength{\belowrulesep}{-0.03pt}
  \begin{adjustbox}{width=\columnwidth, center}
    \begin{tabular}{@{}cccc|cccc@{}}
      \toprule
      \multicolumn{4}{c|}{\textbf{Attack Time Statistics (ms)}}   & \multicolumn{4}{c}{\textbf{Attack Trace Event Statistics (\#)}} \\ \cmidrule(l{\dimexpr0.1cm+1pt} r{\dimexpr0.1cm+1pt}){1-4}\cmidrule(l{\dimexpr0.1cm+1pt} r{\dimexpr0.1cm+1pt}){5-8}
      \multirow{2}{*}{\textbf{\begin{tabular}[c]{@{}c@{}}Exploit\\ DirtyPipe\end{tabular}}} & \multirow{2}{*}{\textbf{\begin{tabular}[c]{@{}c@{}}Escalate\\ privilege\end{tabular}}} & \multirow{2}{*}{\textbf{\begin{tabular}[c]{@{}c@{}}Load \\ .ko\end{tabular}}} & \multirow{2}{*}{\textbf{\begin{tabular}[c]{@{}c@{}}Clear\\ Buffer\end{tabular}}} & \multirow{2}{*}{\textbf{\begin{tabular}[c]{@{}c@{}}\#Total \\ Logs\end{tabular}}} & \multicolumn{3}{c}{\textbf{\#Lost Logs (Min/Avg/Max)}}            \\ \cmidrule(l{\dimexpr0.2cm+2pt} r{\dimexpr0.2cm+2pt}){6-8} & & & & & \textbf{$T_p$: 100\textmu s} & \textbf{1ms} & \textbf{15ms}~\cite{ahmad2022hardlog}      \\ \cmidrule(l{\dimexpr0.1cm+1pt} r{\dimexpr0.1cm+1pt}){1-4}\cmidrule(l{\dimexpr0.1cm+1pt} r{\dimexpr0.1cm+1pt}){5-8}
      3.73 & 3.76 & 4.89 & 0.02 & 1,407 & 0/2/\textbf{7} & 3/16/\textbf{35} & 7/618/\textbf{all} \\ \bottomrule
      \end{tabular}
  \end{adjustbox}
\end{table}

\noindent\textbf{Results.}
\cref{tab:demo_attack} shows the statistics of the attack, indicating an average total attack time of \textms{12.4} with \num{1407} attack-related logs generated.
Threats from semantic bugs (\eg DirtyPipe) can be efficiently exploited by attackers (as stated in \cref{subsec:limitations_of_existing_protection_solutions} and in~\cref{tab:demo_attack}), giving them a high chance of removing considerable attack traces before a single large protection window (\eg removed 618 out of 1,407 logs on average given a \textms{15} protection window), or compromising full attack traces.
Reducing the protection window significantly, by 93.3\% to \textms{1} or by 99.3\% to \textus{100}, proves effective in hindering log tampering.
In these scenarios, the attacker can delete only 3\% (35 logs) or 0.2\% (7 logs) of total logs at most, respectively.

Notably, as the last step for log deletion, installing a malicious kernel module requires around \textms{4.9}.
This time span significantly exceeds given a $T_p$ of \textms{1}.
As a result, all logs during the exploit of DirtyPipe and escalating privilege were logged and protected, even with a $1ms$ $T_p$. In fact, only the ``final millisecond'' logs during module installation were lost.
Hence, this deadline (of \textms{1}) setting also preserved the availability of the full attack trace.
\PP{Memory corruption attack discussion.}
Admittedly, a pure memory corruption attack like in-kernel ROP can {\em avoid kernel module loading} after privilege escalation and tamper with logs faster, by memory write gadgets.
However, unlike userspace exploits, chaining gadgets in-kernel for arbitrary code execution may be challenging and time-consuming.
Our reproduced real-world memory corruption exploits (see Appendix~\cref{appen:all_attacks}) also demonstrate a lengthy attack preparation of more than 1$s$, making it impossible to use such attacks to compromise logs within \textms{1} windows.
%
This is due to the time-consuming nature of manipulating memory layout (\eg heap spray), lack of fault tolerance for attack debugging, and defenses like data execution prevention (DEP) and guarded control stacks (GCS)~\cite{gcs}.
%
So far, loading a kernel module remains a popular rootkit~\cite{rootkit,rootkit-log} to disable logging.

Regardless, when shrinking the protection window to a real-time value of around \textms{1}, \codename significantly raises the difficulty for adversaries to alter the vast majority of their attack traces.
%

%



\subsection{TCB and Vulnerability Discussion}
\label{subsec:tcb_evaluation}

\codename's TCB is comprised of \sm and \hd.
\sm requires an additional 2.2k LoC based on TF-A (29k LoC\footnote{
%
The LoC of TF-A is calculated by counting the compiled sources. 
All uncompiled other platform or SoC-related driver and library sources are excluded.
}),
while the \hd requires 0.8k LoC.
By debloating the driver via record-and-replay, it avoids the adoption of a commodity Linux driver that contains more than 10k LoC and complex kernel dependencies~\cite{guo2022driverlet}.
In total, \codename requires a TCB of around 32k LoC.
This is significantly smaller than typical virtualization software (\eg larger than 862k LoC with KVM~\cite{hof2022blackbox}), or the TZ components (\eg 300k LoC of OP-TEE) used by existing solutions~\cite{varun2023omnilog,shepherd2017emlog}.
Moreover, unlike prior work, \codename does not rely on full-fledged components (both hypervisor and TrustZone), thereby avoiding sharing with their interfaces and attack surfaces (\cref{tab:general_purpose_attack_surface}).
%
%
We further analyzed the security properties of \codename in terms of attack surface reduction (compared to existing solution~\cite{varun2023omnilog}) in Appendix \cref{appen:vul_mitigation}.
In conclusion, \codename reduced their TCB by $9.4-26.9\times$, and avoided all the general-purpose attack surfaces.
\section{Performance Evaluation}
\label{sec:performance_evaluation}

All experiments were performed on our JUNO R2 computer~(\cref{sec:impl}).
For client-server programs, the client workload operated on a desktop with Intel(R) Core(TM) i7-10700 CPU, 16GB RAM, and a 512GB hard disk, connected to the JUNO server via Gigabit Ethernet.

Unless otherwise specified, we measured \codename's performance with the $T_p$ of \textms{1}, since it can prevent attack trace tampering even in strong attacks~(\cref{subsec:case_study}).
We configured \codename's kernel buffer pool with 16 buffers (a total of 1MB size).
This setting very rarely experienced buffer fill-up delays (\cref{subsec:design_realtime_deadline}).

\begin{table}[t]
  \caption{Real-world application and workload description.}
  \vspace{-0.2cm}
  \label{tab:workloads}
  \setlength{\aboverulesep}{0pt}
  \setlength{\belowrulesep}{0pt}
  \begin{adjustbox}{width=\columnwidth, center}
  \begin{tabular}{@{}l|l@{}}
  \specialrule{0.1em}{\aboverulesep}{0pt}
  \textbf{Application}       & \textbf{Workload Description}                                                                                                                                                 \\ 
  \specialrule{0.05em}{0pt}{\belowrulesep}
  \specialrule{0.05em}{1pt}{\belowrulesep}
  \multirow{2}{*}{Nginx}     & \multirow{2}{*}{\begin{tabular}[c]{@{}l@{}}Default 4 worker threads; tested with the \texttt{ApacheBench} \\ of 10K requests for a default file and 32 client concurrency.\end{tabular}}   \\
                             &                                                                                                                                                                               \\ \midrule
  \multirow{2}{*}{Apache}    & \multirow{2}{*}{\begin{tabular}[c]{@{}l@{}}Default settings; tested with the \texttt{ApacheBench} of \\ 10K requests for a default file and 32 client concurrency.\end{tabular}}           \\
                             &                                                                                                                                                                               \\ \midrule
  \multirow{2}{*}{Redis}     & \multirow{2}{*}{\begin{tabular}[c]{@{}l@{}}Default 16 databases; tested by \texttt{memtier\_benchmark} of 1M \\ sets and 1M gets for 32 bytes data with 50 clients.\end{tabular}}      \\
                             &                                                                                                                                                                               \\ \midrule
  \multirow{2}{*}{Memcached} & \multirow{2}{*}{\begin{tabular}[c]{@{}l@{}}Default settings; tested by \texttt{memtier\_benchmark} of 1M set\\ and 1M get for 32 bytes data with 50 clients.\end{tabular}}         \\
                             &                                                                                                                                                                               \\ \midrule
  \multirow{2}{*}{MySQL}     & \multirow{2}{*}{\begin{tabular}[c]{@{}l@{}}Default 10 tables with table size 10,000; tested by \texttt{sysbench} \\ \texttt{oltp\_read\_write} for total 10K transactions ($\geq$ 200K queries).\end{tabular}} \\
                             &                                                                                                                                                                               \\ \midrule
  7zip                       & Phoronix benchmark \texttt{pts/compress-7zip}.                  \\ \midrule
  OpenSSL                    & Phoronix benchmark \texttt{pts/openssl}.                        \\ \midrule
  Firefox                    & \texttt{Speedometer 2.0} benchmark.                             \\ \midrule
  GNU Octave                 & Phoronix benchmark \texttt{system/octave-benchmark}.            \\ \midrule
  Wget                       & No-cache and quiet mode; fetching a default file with 10K runs.     \\ 
  \specialrule{0.1em}{\aboverulesep}{0pt}
  \end{tabular}
  \end{adjustbox}
  \vspace{-0.3cm}
\end{table}

\PP{Comparison settings.}
We used four comparison settings to evaluate our system, which we describe in the following paragraphs.

\textbf{Native} indicates system performance without any kind of logging.
\textbf{Native-OBS} is the performance when logs are captured and then in memory by the native observability generator~(extended Tracee~\cite{tracee}).
Native-OBS does not persist or protect logs, thus serving as the ideal (maximum) performance in our system with logging.

To show the advantage of hardware permission-based protection,
we also evaluated a \codename variant employing \textit{EL3-memcpy}-based~(\cref{subsec:overview_inmemory_observability_protection}) protection (labeled \textbf{\codename-MCPY}), which copies logs from the OS domain to the protected.
Furthermore, to show the cost of storage operations within our system, we also evaluated a \codename variant without any such operations, called \textbf{\codename-IM}.
We evaluate \textbf{\codename-S2PT} and \textbf{\codename-GPT}, by using S2PT and GPT, respectively.

Finally, we compared \codename with the state-of-the-art synchronous protection approach, OmniLog~\cite{varun2023omnilog} (referred to as {\bf \omnilog}).
Note that OmniLog is designed for audit (system) log protection and integrated with Linux Auditd~\cite{auditd}.
%
For a fair comparison, we made our best efforts to reproduce OmniLog and incorporated it with our observability generator, focusing only on in-memory protection (\ie logs are discarded after synchronously being copied into EL3 memory by \omnilog).
This avoids storage overhead to optimistically approximate its performance.

\begin{table}[]
  \caption{Throughput under the {\em getpid-flood}~(\cref{subsec:protection_scheme_evaluation}) benchmark.
  }
  \vspace{-0.2cm}
  \label{tab:throughput_pidloop}
\setlength{\aboverulesep}{-0.05pt}
  \setlength{\belowrulesep}{-0.05pt}
  \setlength{\tabcolsep}{2pt}
\begin{adjustbox}{width=0.9\columnwidth, center}
\begin{tabular}{@{}|l|c|c|@{}}
\toprule
\textbf{Approach}                                            & \textbf{Throughput (logs/sec)}         & \textbf{Relative Percent (\%)} \\ \midrule
{\bf Native-OBS}    & 201,038      & 100\%    \\ \midrule
\begin{tabular}[c]{@{}l@{}}{\bf \codename-IM}\\ $T_p$: {\textms{1}\big/\textus{500}\big/\textus{100}}\end{tabular}   & 178,571\bigg/172,053\bigg/166,500 & 88.8\%\bigg/85.5\%\bigg/81.3\%  \\ \midrule
\begin{tabular}[c]{@{}l@{}}{\bf \codename-GPT}\\ $T_p$: {\textms{1}\big/\textus{500}\big/\textus{100}}\end{tabular}      & 177,399\bigg/168,711\bigg/154,966 & 88.2\%\bigg/83.9\%\bigg/77.1\%  \\ \midrule
\begin{tabular}[c]{@{}l@{}}{\bf \codename-S2PT}\\ $T_p$: {\textms{1}\big/\textus{500}\big/\textus{100}}\end{tabular}      & 173,094\bigg/165,253\bigg/155,603 & 86.1\%\bigg/82.2\%\bigg/77.4\%  \\ \midrule
\begin{tabular}[c]{@{}l@{}}{\bf \codename-MCPY} \\ $T_p$: {\textms{1}\big/\textus{500}\big/\textus{100}}\end{tabular} & 150,466\bigg/137,757\bigg/120,376 & 74.8\%\bigg/68.5\%\bigg/59.9\%  \\ \midrule
 {\bf \omnilog}       & 99,143                      & 49.5\%               \\ \bottomrule
\end{tabular}
\end{adjustbox}
\vspace{-0.2cm}
\end{table}

\begin{figure}[t]
  \centering
  \includegraphics[width=\linewidth]{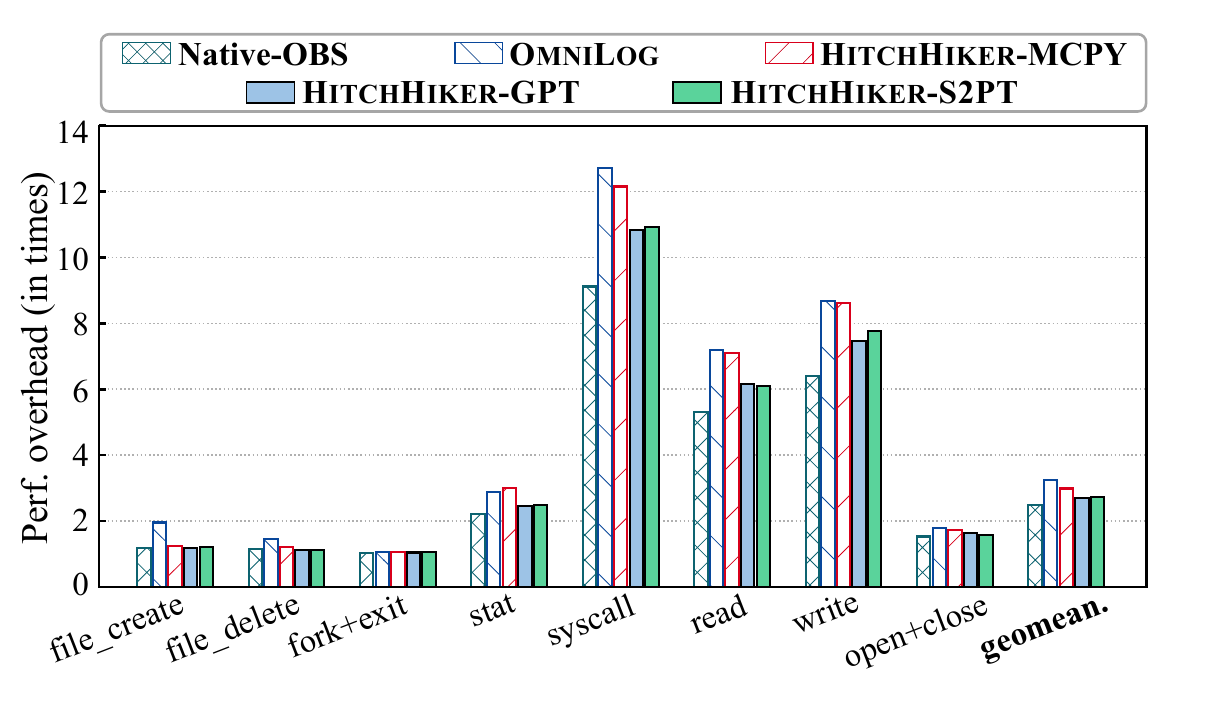}
  \vspace{-0.7cm}
  \caption{Overhead of \codename on LMBench~\cite{larry1996lmbench}.}
  \label{fig:overhead_lmbench}
  \vspace{-0.4cm}
\end{figure}

\subsection{Micro-benchmarks}
\label{subsec:micro_benchmarks}

This section describes \codename's throughput breakdown and its system-event-level performance.

\PPn{Log throughput breakdown with {\em getpid-flood}}. 
To break down the impact of different aspects of \codename (\eg protection, persistence), we leveraged the \textit{getpid-flood} micro-benchmark (\cref{subsec:protection_scheme_evaluation}) to generate logs intensively evaluate the performance.
%

%

\vspace{0.2em}
\noindent\textit{Results.}
\cref{tab:throughput_pidloop} illustrates the results.
The throughput of Native-OBS (201,038 logs/second) acts as the reference point, representing the theoretical maximum observability log throughput of our system.
\codename-GPT reaches a throughput of 177,399 entries per second ($88.2\%$ of Native-OBS) under the \textms{1} deadline setting.
Similarly, at such settings, \codename-S2PT has $86.1\%$ of the Native-OBS throughput.
Furthermore, as the timer frequency $T_p$ shortens from \textms{1} to \textus{100}, there is a corresponding decline in the relative throughput of \codename (from $88.2\%$ to $77.1\%$).
This is expected, as more frequent protection enforcement introduces more system overheads for \codename to respond.
%

\codename-MCPY has a throughput varies between $74.8\%$ and $59.9\%$, according to the $T_p$ setting ranging from \textms{1} to \textus{100}. 
This further shows the advantage of the hardware permission switch over memory copy.
\omnilog displays a relative throughput of $49.5\%$, underscoring its significant slowdown due to frequent synchronous system blocks and log protections.

%
Compared to \codename-IM, which solely employs memory protection, \codename (S2PT and GPT) experiences an additional $0.6\%$ to $4.2\%$ reduction in throughput depending on the deadline settings.
This overhead is attributed to the activation of \hd's manager thread, which causes extra system burdens (\eg task synchronization and I/O).
Nevertheless, the performance impact of the background log persistence remains limited, primarily due to the utilization of high-bandwidth SATA SSD.

\PP{System event latency}
We evaluated \codename's slowdown upon kernel operations by using the widely adopted~\cite{paccagnella2020custos,ahmad2022hardlog} {\em lmbench}~\cite{larry1996lmbench}.
We executed latency benchmarks for (a) file system creation/deletion, (b) process creation/exit, and (c) syscall.

\vspace{0.1em}
\noindent\textit{Results.}
\cref{fig:overhead_lmbench} shows the evaluation results.
\codename-GPT and \codename-S2PT imposes a geometric mean of $1.69\times$ and $1.71\times$ overhead more than Native, while adding only $8.4\%$ and $9.3\%$ overhead to Native-OBS~($1.48\times$ more than native), respectively.
The overheads for the memory copy-based approaches, \codename-MCPY and \omnilog, are respectively $1.99\times$ ($20.6\%$) and $2.25\times$ ($31\%$) higher than Native-OBS.
As the lmbench only generates kernel events, the large overheads for \codename mainly come from the Native-OBS ($1.48\times$ more than Native).
This overhead is amortized during real-world workload execution~(next section).
In the future, observability generation overhead can be optimized using eBPF compile-optimization~\cite{mao2024merlin} and efficient log collection~\cite{lim2021provbpf,sekar2024eaudit}.

\subsection{Real-world Programs}
\label{subsec:performance_evaluation}

This section describes \codename's performance on real-world workloads using common benchmarks.
The overhead is compared with different system settings mentioned in~\cref{sec:performance_evaluation}.

\PP{Settings.}
We chose five common client-side applications (7zip, OpenSSL, Firefox, GNU Octave, and Wget) and server-side applications (Nginx, Apache, Redis, Memcached, and MySQL). 
We divide applications by client and server workloads, because the latter typically produce significantly higher amounts of logs.
These applications have also been evaluated under similar workloads by prior work~\cite{ahmad2022hardlog,varun2023omnilog,riccardo2020kennylogging}.
Detailed workloads are described in \cref{tab:workloads}. 

\begin{figure}[t]
  \centering
  \includegraphics[scale=0.42]{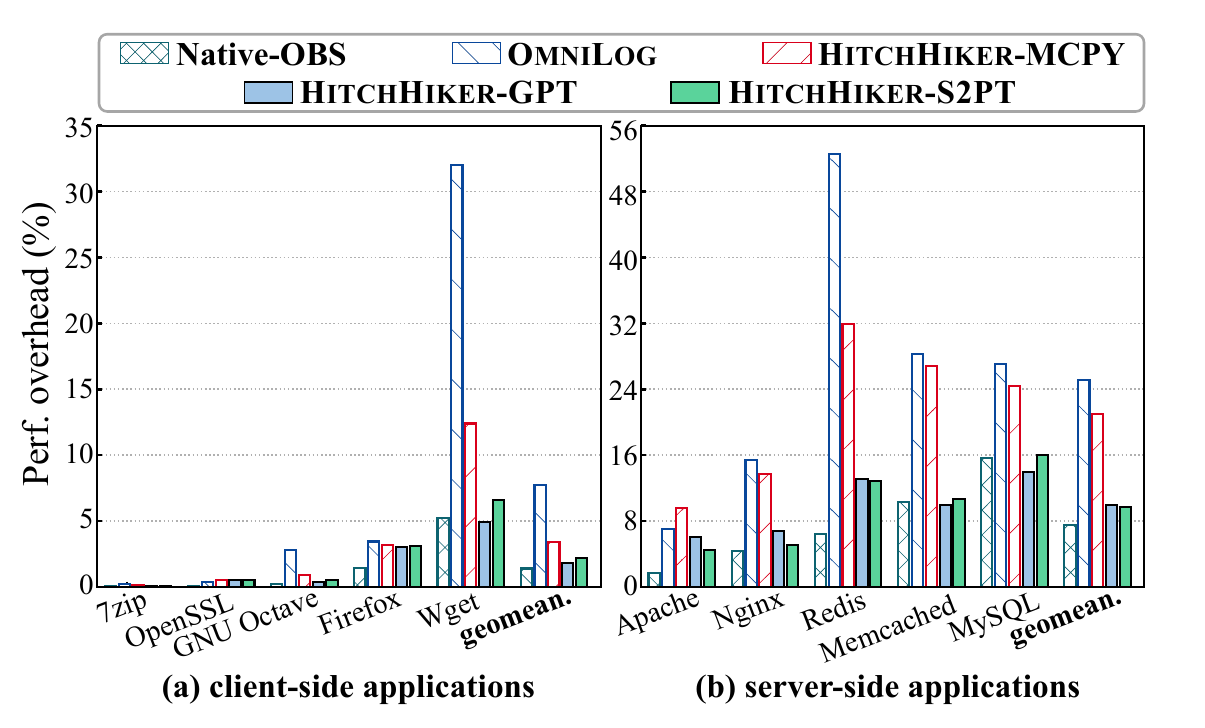}
  \vspace{-0.7cm}
  \caption{Real-world workload overheads.
  From left to right, the log throughput for each program is: 1907, 589, 32358, 7234, 31434; 10999, 78776, 78537, 93928, 101962 logs/sec.
  %
  }
  \label{fig:overhead_realworld}
  \vspace{-0.5cm}
\end{figure}

\PP{Results.}
\cref{fig:overhead_realworld} illustrates the performance overhead incurred by real-world applications.
\codename-GPT exhibits a geometric mean overhead of $1.8\%$ and $9.9\%$ over Native in client-side and server-side programs, respectively.
Similarly, \codename-S2PT's geometric mean overheads are $2.2\%$ and $9.7\%$.
In contrast, \codename-MCPY's geometric mean overheads are $3.4\%$ and $20.9\%$ over Native, on client-side and server-side programs, respectively.
In the worst-case, \codename-MCPY's overhead is up to $31.9\%$, while \codename's worst remains at only $13.9\%$.
Recall the protection cost described in~\cref{tab:prot_primitives}; \textit{el3-memcpy} is expensive due to its CPU-intensive nature and its lack of optimizations in EL3 software~\cite{el3-memcpy}.
Thus, \codename-MCPY imposes an overhead of $0.5-1.2\times$ more than \codename-GPT and \codename-S2PT, which leverage efficient hardware permission switch~(\cref{subsec:design_realtime_deadline}).
\omnilog also adopts {\em el3-memcpy} for in-memory log protection, but also {\em synchronously} protects every log.
%
It results in a geometric mean of $8\%$ client-side and $26\%$ server-side overheads over Native.
%
%
Client programs typically focus on computational tasks (\eg compression, encryption, and calculation) and thus are ``{\em log-sparse}'' that generate fewer logs. 
However, Wget (a web downloader) is an exceptional client program.
Due to its frequent system calls and network transitions, \omnilog imposes a $30\%$ overhead over Native on it.
Conversely, server-side applications are characterized as ``{\em log-intensive}'', owing to their frequent I/O tasks for invoking system calls, recording application logs into local file systems, and communicating with clients through the network.
Thus, given the high observability log throughput, \omnilog results in a large overhead (\eg up to $52\%$ over Native in Redis) due to intensively blocking the system execution to wait for log protection. 
In all scenarios, \omnilog's synchronous protection incurs $0.12-0.65\times$ overheads over
asynchronous memory copying (\codename-MCPY), or $0.2-3\times$ over hardware-based permission switching (\codename).



\PP{Takeaway.}
Despite enforcing short real-time deadlines (around \textms{1}), \codename provides high performance.
On {\em log-sparse} workloads, all solutions provide decent performance (less than $8\%$ overhead), yet \codename provides near-native performance ($2\%$ geometric mean overhead only).
On {\em log-intensive} applications, synchronous or periodic copying of logs to privileged memory becomes inefficient
(up to $52.6\%$ overheads).
Here, \codename with permission switch incurs a geometric mean of only $9.9\%$ (up to $15.9\%$ only), which is up to 77.5\% and 61.9\% lower than OmniLog on client and server-side programs, respectively, showing its efficiency.
%


%

\vspace{-0.2cm}
\section{Timer Protection Discussion}
\label{sec:timer-discussion}

A limitation of our current implementation~(\cref{sec:impl}) is that attackers may use kernel vulnerabilities in the kernel, even before it is compromised, to trick it into manipulating the GIC register; thereby disabling timer interrupt prioritization.
For instance, attackers may leverage code reuse (e.g., ROP) attacks.
We found that the preparation steps for such attacks typically take a long time~($\geq$ \textms{100}, Appendix \cref{appen:all_attacks}); thus, the initial logs (before GIC manipulation) will still be protected to reveal attack preparations.
Nevertheless, such an attack would give attackers an extended window to tamper with the remaining logs during full system compromise.

There are two potential solutions to prevent the aforementioned attack.
The first solution is to write-protect the MMIO interface of GIC's interrupt priority register (e.g., using S2PT or GPT to remove the permissions for accessing the MMIO page in the OS domain).
By doing so, only the security monitor controls the interface of GIC interrupt priority registers; thus, any attempts to manipulate them are interposed by the monitor.
The second solution is to isolate the entire timer interrupt handling into the protected domain.
In particular, the monitor (\sm) may (a) initialize the timer interrupt and lock it down to a secure interrupt (in GICv2~\cite{wang2022rttee}), or (b) set the interrupt route model to route it exclusively to EL3 (in GICv3~\cite{irqroute}).
We leave implementing these solutions to future work.

\section{Related Work}
\label{sec:related}
\noindent{\bf System Observability.}
Observability is crucial for system dissection and analysis with a broad spectrum of applications~\cite{li2022enjoy, minna2023sok}.
Researchers utilize different levels of observability by capturing program semantics via application logs~\cite{he2021asurvey} for anomaly detection~\cite{du2017deeplog,zhang2019robust} and failure diagnosis~\cite{yu2014comprenhending,cotroneo2019howbad}; 
monitoring syscall-level audit logs~\cite{inam2023history} for attack detection~\cite{han2020unicorn,zeng2022shadewatcher} and investigation~\cite{king2003backtracking,xu2022depcomm}; 
and utilizing network logs~\cite{fonseca2007xtrace} for malware detection~\cite{fu2021realtime,fu2022encrypted}, forensics and traffic classification~\cite{zhou2011securenet,ujcich2021causal}.
Fusing multi-layer observability integrates a holistic security profile.
This is achieved by incorporating application logs~\cite{hassan2020omegalog,yu2021alchemist}, network logs~\cite{datta2022alastor,adam2017transparent}, library function logs~\cite{wang2018lprov}, and instruction traces~\cite{PalanTir22} with audit logs, strengthens system forensics and supports fine-grained provenance reconstruction.
\codename is the first system for cross-layer system observability protection and maintenance.

\PP{Trusted execution environment.}
Enclaves~\cite{cc}
enable applications like stream processing~\cite{park2019streamboxtz}, mobile app protection~\cite{santos2014using}, control flow attestation~\cite{abera2016cflat}, security policy enforcement~\cite{brasser2016regulating}, and serverless computing~\cite{zhao2023reusable,li2021pie}.
Researchers explored building enclaves with different abstraction levels, including userspace process~\cite{chen2008overshadow,hofmann2013inktag}, secure containers~\cite{arnautov2016scone,shen2019xcontainers}, and confidential VMs~\cite{hua2017vtz,li2021twinvisor}.
Works adopt hardware protections like TZASC~\cite{brasser2019sanctuary}, S2PT~\cite{hof2022blackbox,han2023mytee}, GPT~\cite{zhang2023shelter,sridhara2023acai}, or even hardware-software co-design~\cite{feng2021penglai,du2023accelerating} to support domain enforcement for general-purpose enclaves.
Inspired by those works, \codename employs hardware features to tailor the secure environment exclusively for its secure log management.

\PP{Tamper-evident logging.}
Instead of providing log integrity and availability against tampering attacks, \textit{tamper-evident} schemes detect tampering using cryptographic integrity proofs.
Logs are either asynchronously~\cite{karande2017sgxlog,paccagnella2020custos,bowers2014pillarbox,shepherd2017emlog} or synchronously~\cite{riccardo2020kennylogging,hoang2022faster} signed by Message Authentication Code (MAC).
After creating signatures, logs as well as their signatures (as the proof-of-integrity) are securely sealed (encrypted) on local storage and later sent to the central servers for further authentication.
%
%
%
In the future, combining tamper-evident logging and \codename can lead to the best of both worlds, namely the real-time availability and tamper evidence.





\section{Conclusion}
\label{sec:conclusion}

%

\codename is an efficient and high-assurance observability protection system.
It leverages efficient memory permission switches and hardware timers to enforce real-time short and configurable log protection deadlines.
%
By its first principles approach of secure environment design, it debloats the secure environment and avoids sharing attack surfaces with other system components.
Compared to prior work, \codename achieves $93.3-99.3\%$ shorter log protection deadlines (protecting the vast majority of logs), $9.4-26.9\times$ smaller TCB, while reducing performance overheads by $61.9-77.5\%$.
%

\begin{acks}
We thank the shepherd and anonymous reviewers for their insightful feedback in finalizing this paper.
We also thank Michael Swift for his comments on the manuscript and Jinting Wu from COMPASS Lab for his assistance with the experiments.
This research/project is supported by the National Research Foundation, Singapore under its Industry Alignment Fund – Pre-positioning (IAF-PP) Funding Initiative, the US Air Force Office of Scientific Research (AFOSR) under award number FA9550-24-1-0204, and the National Natural Science Foundation of China under Grant No.62372218.
Any opinions, findings and conclusions or recommendations expressed in this material are those of the authors and do not reflect the views of National Research Foundation, Singapore.
%
\end{acks}

\bibliographystyle{plain}
\balance
\bibliography{main}

\clearpage
\begin{appendices}

\section{Kernel Exploitations for Log Tampering Attacks}
\label{appen:all_attacks}

\begin{table}[t]
    \centering
    \caption{
    Performed log tampering attacks by exploiting host kernel vulnerabilities.
    } 
    \label{tab:fast_attacks}
    \setlength{\aboverulesep}{0.05pt}
    \setlength{\belowrulesep}{0.05pt}
    \setlength{\tabcolsep}{3pt}
    \renewcommand{\arraystretch}{1.0}
    \begin{adjustbox}{width=1\linewidth}
    \begin{threeparttable}
        \begin{tabular}{@{}llr@{}}
            \toprule
            \textbf{Vuln. CVE} & \textbf{Exploitation Description (Type\tnote{\bf \#}\ \ )}   & \textbf{Time (ms)\tnote{\bf *}} \\ \midrule
            2022-0847       & \begin{tabular}[c]{@{}l@{}}Write arbitrary files via page cache splicing with \\ improper \textcode{PIPE\_BUF\_FLAG\_CAN\_MERGE} flag.\\ {\bf(S: Improper Preservation of Permissions)}\end{tabular}                    & $9.8 - 14.4$            \\ \midrule
            2023-0386       & \begin{tabular}[c]{@{}l@{}}Copy a capable file from a \textcode{nosuid} file mount \\ into another mount due to the uid mapping bug.\\ {\bf(S: Improper Ownership Management)}     \end{tabular}                                               & $12.1 - 15.8$         \\ \midrule
            2023-2640       & \begin{tabular}[c]{@{}l@{}}Set privileged extended attributes on mounted files \\ due to skipped capability check in \textcode{ovl\_copy\_xattr} \\ {\bf(S: Incorrect Authorization)}\end{tabular}   & $13.5 - 25.5$         \\ \midrule
            2023-32629       & \begin{tabular}[c]{@{}l@{}}Extend the mounted file privileges due to skipped \\ capability check in \textcode{ovl\_copy\_up\_meta\_inode\_data} \\ {\bf(S: Incorrect Authorization)}\end{tabular}   & $13.8 - 27.1$         \\ \bottomrule
            2023-2640       & \begin{tabular}[c]{@{}l@{}}Set privileged extended attributes on mounted files\\ due to lack of security checks in {\em ovl\_do\_setxattr}.\\ {\bf(S: Incorrect Privilege Authorization)} \end{tabular} & $40 - 50$            \\ \midrule
            2020-14386      & \begin{tabular}[c]{@{}l@{}}Overwrite the buffer in a packet of the network \\ namespace due to integer overflow in {\em tpacket\_rcv}.\\ {\bf(M: Out-of-bounds Write)} \end{tabular}                                      & $\ge 100$      \\ \midrule
            2022-0185      & \begin{tabular}[c]{@{}l@{}}Overwrite the heap due to \\ vulnerability in {\em legacy\_parse\_param}.\\ {\bf(M: Heap-based Overflow)} \end{tabular}                                      & $\ge 1000$      \\ \midrule
            2023-32233      & \begin{tabular}[c]{@{}l@{}} Use-after-free in Netfilter \textcode{nf\_tables} \\ when processing batch requests.\\ {\bf(M: Use-After-Free)} \end{tabular}                                      & $\ge 1000$      \\ \midrule
            2023-0179      & \begin{tabular}[c]{@{}l@{}} Add an \textcode{nft\_payload} expression trigger the stack \\ buffer overflow via the \textcode{rule\_add\_payload} \\ {\bf(M: Stack Buffer Overflow)} \end{tabular}                                      & $\ge 1000$     \\ \midrule
            2022-2588    & \begin{tabular}[c]{@{}l@{}}Use after free because of the \\ \textcode{cls\_route} filter implementation \\ {\bf(M: Use-After-Free)} \end{tabular}                                      & $\ge 1000$      \\ \bottomrule
        \end{tabular}
    \begin{tablenotes}
        \item[\textbf{*}]
        {Time includes vulnerability exploitation and malicious module injection for log buffer deletion if the exploitation within \textms{15}, otherwise the PoC execution time. Note: This time span may vary depending on the specific machine and payload.}
        \vspace{0.1cm}
        \item[\textbf{\#}] Type
        {{\bf S:} semantic bugs; {\bf M:} memory corruptions.}
    \end{tablenotes}
    \end{threeparttable}
    \end{adjustbox}
\end{table}

We conduct end-to-end log tampering attacks by leveraging real-world Linux kernel CVEs.
The local attack setup and assumptions are the same as detailed in~\cref{subsec:case_study}.

We reproduced ten CVEs, with five semantic vulnerabilities (bugs) and five memory corruption vulnerabilities.
For the PoCs that can lead to a full log compromise within \textms{15}, we report the total time including executing the PoC exploit and injecting the malicious kernel module, which is the total time of a {\em log tampering attack}.
Otherwise, we simply execute the PoC exploits and report their time (since exploiting the vulnerabilities requires considerably longer than the \textms{15} protection window mentioned in~\cref{subsec:limitations_of_existing_protection_solutions}).
All statistics are reported in \cref{tab:fast_attacks}.

Our experimental results show that exploiting memory corruption bugs requires a long preparation time ($\geq$ \textms{100}) while exploiting semantic bugs can be brutal.
To exploit memory corruption bugs, the adversary has to carefully manipulate kernel objects (e.g., heap spraying)~\cite{lin2022dirtycred} and then bypass defense mechanisms like KASLR.
Those attack preparations are time-consuming.






\begin{table}[th]
  \caption{
  OmniLog~\cite{varun2023omnilog}'s TCB (LoC) breakdown.
  }
  \vspace{-0.2cm}
  \label{tab:omnilog-tcb}
  \setlength{\tabcolsep}{3pt}
  \setlength{\aboverulesep}{-0.03pt}
  \setlength{\belowrulesep}{-0.03pt}
  \begin{adjustbox}{width=\linewidth,center}
\begin{tabular}{@{}ccc|ccc@{}}
\toprule
\multicolumn{3}{c|}{\textbf{TrustZone solution}} & \multicolumn{3}{c}{\textbf{Hypervisor solution}} \\ \midrule
\begin{tabular}[c]{@{}c@{}}Monitor\\ (TF-A)\end{tabular} & \begin{tabular}[c]{@{}c@{}}Trusted kernel\\ (OP-TEE)\end{tabular} & Total & \begin{tabular}[c]{@{}c@{}}Monitor\\ (KVM)\end{tabular} & \begin{tabular}[c]{@{}c@{}}Trusted kernel\\ (Linux)\end{tabular} & Total \\\midrule
$\sim$29k + 0.5k & $\sim$300k + 0.1k & \textgreater{}329.6k & $\sim$864k + 0.5k & $\sim$27.8M + 0.3k & \textgreater 27.8M \\ \bottomrule
\end{tabular}
  \end{adjustbox}
\end{table}

\begin{table*}[t]
  \caption{
  TrustZone issues (from~\cite{cerdeira2020tzsok}) and \codename's mitigations.
  }
  \vspace{-0.2cm}
  \label{tab:tzsok}
  \setlength{\tabcolsep}{3pt}
  \setlength{\aboverulesep}{-0.03pt}
  \setlength{\belowrulesep}{-0.03pt}
  \begin{adjustbox}{width=\textwidth,center}
\begin{tabular}{|c|l|l|}
\hline
\multicolumn{1}{|l|}{\textbf{Category}} & \textbf{Issue} & \textbf{Mitigation} \\ \hline
\multirow{9}{*}{\textbf{\begin{tabular}[c]{@{}c@{}}Arch.\\ issues\end{tabular}}} & I01. SW drivers run in the TEE kernel space & \multirow{3}{*}{\begin{tabular}[c]{@{}l@{}}\codename does not use trusted (TEE) kernels or TrustZone secure world.\\ \codename's secure environment is built upon hardware protection primitives\\ with its debloated software stack (\cref{subsec:design_security_monitor}).\end{tabular}} \\ \cline{2-2}
 & I02. Wide interfaces between TEE system subcomponents &  \\ \cline{2-2}
 & I03. Excessively large TEE TCBs &  \\ \cline{2-3} 
 & I04. TAs can map physical memory in the NW & \multirow{6}{*}{\begin{tabular}[c]{@{}l@{}}\codename does rely on TAs, but its native protected log daemon \hd (\cref{subsec:design_daemon}).\\ 
 \hd is exclusively programmed and provisioned by trustworthy enterprise IT\\ 
 administrators. The integrity of \hd is attested by the security monitor \sm \\
 during collaborative loading (\cref{subsubsec:daemon_execution}). The communication interfaces of \hd is \\
 exclusively  controlled by administrators with established secure channel (\cref{subsubsec:design_daemon_secret}) \\ 
 and sanitized network communication (\cref{subsubsec:design_daemon_syscall}).\end{tabular}} \\ \cline{2-2}
 & I05. Information leaks to NW through debugging channel &  \\ \cline{2-2}
 & I06. Absent or weak ASLR implementations &  \\ \cline{2-2}
 & I07. No stack cookies, guard pages, or execution protection &  \\ \cline{2-2}
 & I08. Lack of software-independent TEE integrity reporting &  \\ \cline{2-2}
 & I09. Ill-supported TA revocation &  \\ \hline
\multirow{4}{*}{\textbf{\begin{tabular}[c]{@{}c@{}}Input \\ validation \\ issues\end{tabular}}} & I10. Validation bugs within the secure monitor & Minimally exposed security monitor call interface with validations (see \cref{tab:smc}). \\ \cline{2-3} 
 & I11. Validation bugs within TAs & \multirow{2}{*}{\begin{tabular}[c]{@{}l@{}}\codename does not leverage TA or trusted kernels.\end{tabular}} \\ \cline{2-2}
 & I12. Validation bugs within the trusted kernel &  \\ \cline{2-3} 
 & I13. Validation bugs in secure boot loader & Vulnerable (Out of scope). \\ \hline
\multirow{5}{*}{\textbf{\begin{tabular}[c]{@{}c@{}}Functional \\ \&\\ extrinsic\\  issues\end{tabular}}} & I14. Bugs in memory protection & \multirow{4}{*}{\begin{tabular}[c]{@{}l@{}} \codename's protection primitives (both memory and peripheral device) do \\not rely on TrustZone. Configurations of TrustZone (e.g., TZASC or TZPC) will \\ not affect \codename's protected domain which is exclusively controlled by \sm \\ 
in the EL3 root world.\end{tabular}} \\ \cline{2-2}
 & I15. Bugs in conﬁguration of peripherals &  \\ \cline{2-2}
 & I16. Bugs in security mechanisms &  \\ \cline{2-2}
 & I17. Concurrency bugs (from multiple TAs) &  \\ \cline{2-3} 
 & I18. Software side-channels & Vulnerable (Out of scope). \\ \hline
\end{tabular}
  \end{adjustbox}
\end{table*}


\section{Security Property Analysis}
\label{appen:vul_mitigation}

%
This section compares the security properties of \codename against a state-of-the-art in-memory secure environment-based log protection solution, OmniLog~\cite{varun2023omnilog}. 
In particular, we first compare based on TCB size~(\cref{subsec:tcb-omnilog}). 
Then, to further analyze \codename's security property against the general-purpose environment (TrustZone), we describe how \codename tackles the architectural and implementation issues of TrustZone (\cref{subsec:mitigation}).
%
%

\subsection{TCB comparison}
\label{subsec:tcb-omnilog}

%
OmniLog's architectural implementation contains a {\em trusted kernel} (i.e., a Linux VM at VMX non-root mode or a TrustZone OP-TEE secure OS) alongside a {\em privileged security monitor} (i.e., hypervisor at EL2 or TrustZone's trusted firmware at EL3).
%

%
\cref{tab:omnilog-tcb} shows TCB breakdown of OmniLog.
In terms of OmniLog's TrustZone implementation, TF-A (29k LoC) with 0.5k LoC extension alongside OP-TEE (300k LoC) with 0.1k extension is required.
In total, 329.6k LoC of TCB is required.
Regarding OmniLog's hypervisor-based approach, while they were implemented based on full-fledged KVM (more than 864k LoC) with a commercial Linux VM (more than 27.8M LoC), they were considered a micro hypervisor.
Under such settings, a typical micro hypervisor like Bareflank~\cite{Bareflank} requires around 108k LoC.
It also requires a VM with Linux.
In contrast, \codename requires a total of around 32k LoC of TCB (\cref{subsec:tcb_evaluation}).
\codename's TCB reduction factor is more than 10$\times$ for each solution of OmniLog.


\subsection{Vulnerability Mitigation Analysis}
\label{subsec:mitigation}



%
This section describes how \codename tackles the architectural and implementation issues of TrustZone.
We leveraged the Systemization of Knowledge study by Cerdeira et al.~\cite{cerdeira2020tzsok} to categorize the issues of TrustZone into (a) architectural issues, (b) input validation bugs of implementation, and (c) remnant functional and extrinsic implementation bugs.
\cref{tab:tzsok} describes the categories and detailed issues and explains how \codename addresses these issues.

As \codename's secure log environment is built upon hardware primitives (\cref{subsec:overview_secure_environment}) with its debloated software stack (\cref{subsec:design_security_monitor}), it does not rely on TrustZone secure world or the trusted kernel (e.g., OPTEE).
Therefore, the codebase of secure kernels or their complex interfaces will not affect \codename ({\bf I01}-{\bf I03}).
Moreover, protection primitives are controlled exclusively by \sm in EL3 (root world) without sharing TrustZone's memory or peripheral protection ({\bf I15}-{\bf I17}). 
Similarly, \codename does not rely on general-purpose TAs but on its dedicated protected daemon \hd, thereby avoiding TA-related issues ({\bf I04}-{\bf I09}).
In terms of input validation issues, the vast majority of issues (119 out of 121 found issues) were from TAs or the secure kernel ({\bf I11, I12})~\cite{cerdeira2020tzsok}. 
These issues were pruned by \codename. 
For security monitor-related issues, \codename only involves a small set of 8 input interfaces with input validation, as illustrated in (as illustrated in \cref{tab:smc}).
%



%
Out of all potential issues, \codename is still vulnerable to validation bugs in the boot loader ({\bf I13}) and side-channels~({\bf I18}), which are out-of-scope and left for future work. 
Nevertheless, this is a significantly smaller set of issues than prior work like OmniLog.

\chuqic{
%
%
%
%
}







\begin{table*}[th]
  \caption{
  \codename's security monitor call (SMC) interfaces and security validation.
  }
  \vspace{-0.2cm}
  \label{tab:smc}
  \setlength{\tabcolsep}{3pt}
  \setlength{\aboverulesep}{-0.03pt}
  \setlength{\belowrulesep}{-0.03pt}
  \begin{adjustbox}{width=\textwidth,center}
\begin{tabular}{|l|l|l|}
\hline
\textbf{SMC} & \textbf{Description} & \textbf{Callsite and input validation} \\ \hline
secure\_buffer & Protect a log buffer by switching its memory permission. & The buffer address is inside the pre-allocated log buffer pool. \\ \hline
return\_buffer & Return a log buffer from \hd and refill it to the OS's pool. & \begin{tabular}[c]{@{}l@{}}(1) Only called by \hd: the SMC parameter is a communication\\ buffer pointer inside \hd's protected memory.\\ (2) The buffer address is inside the pre-allocated log buffer pool.\end{tabular} \\ \hline
init\_\hd & Initialize \hd's address space and task structures. & The load of \hd binary and its memory is verified and protected. \\ \hline
request\_OS & Request the OS service from \hd's exception vector table. & Only called by \hd in the protected domain. \\ \hline
sched\_in\_\hd & Schedule \hd from the OS. & No parameters. \\ \hline
init\_secureIO & Initialize the protected device driver instance. & Only called during system initialization. \\ \hline
issue\_secureIO & Issue a secure I/O task to persist or retrieve logs. & \begin{tabular}[c]{@{}l@{}}(1) Only called by \hd: the SMC parameter is a communication\\ buffer pointer inside \hd's protected memory.\\ (2) The DMA buffer is inside the pre-allocated DMA range.\end{tabular} \\ \hline
destroy\_\hd & Destroy \hd and persist all remaining protected logs. & \begin{tabular}[c]{@{}l@{}}Only called by \hd: the SMC parameter is a communication\\ buffer pointer inside \hd's protected memory.\end{tabular} \\ \hline
\end{tabular}
\end{adjustbox}
\end{table*}


\section{Protection Deadline Comparison}
\label{subsec:trade-off}

\begin{figure}[t]
  \centering
  \includegraphics[scale=0.42]{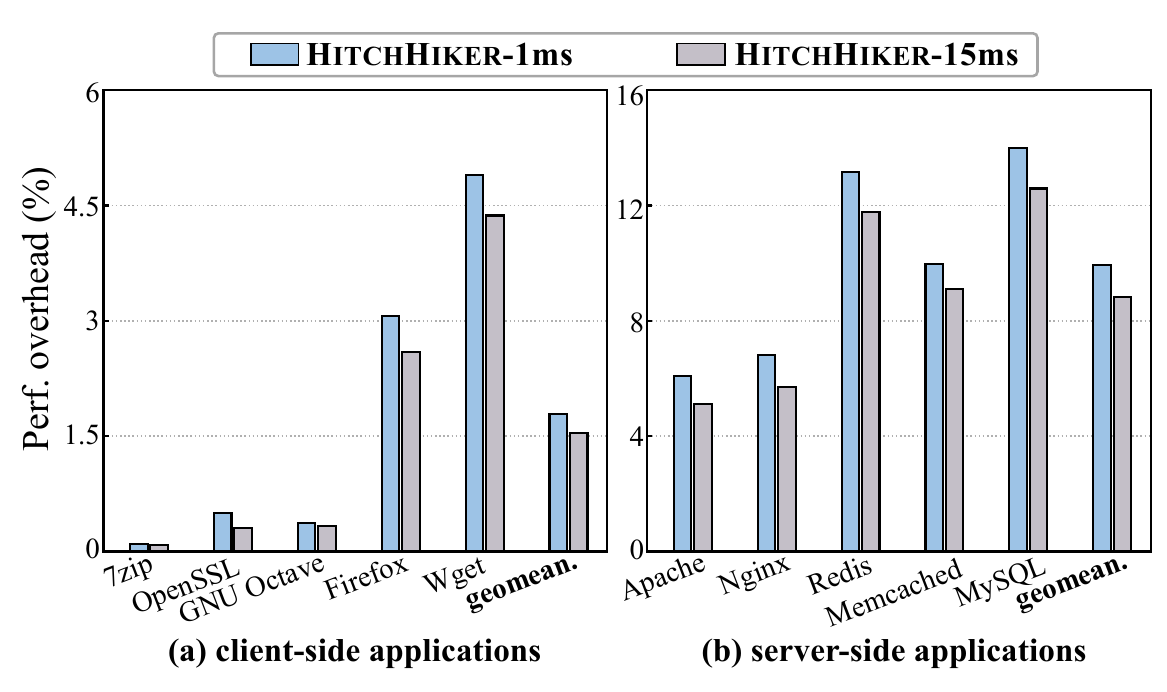}
  \vspace{-0.3cm}
  \caption{
  Real-world workload overheads on different protection deadline settings.
  }
  \label{fig:overhead_tradeoff}
  \vspace{-0.3cm}
\end{figure}

%
While short protection deadlines ($T_p$) settings provide high security by raising the bar against log tampering attacks, they may impact system performance. 
To understand the performance and security trade-off, this section describes \codename's performance on real-world workloads under two different deadline settings.

\PPn{
Settings.
}
%
We configured \codename with two different protection deadline ($T_p$) settings: \textms{1}, which has been proven to be secure to preserve attack traces, and \textms{15}, which may leave the window for attackers to clear the traces (\cref{subsec:case_study}).
We evaluated \codename-1ms and \codename-15ms with the same real-world programs and workloads described in \cref{tab:workloads} with GPT deployed.

\PPn{
Results.
}
%
\cref{fig:overhead_tradeoff} shows the overhead under different protection deadline settings.
\codename-1ms incurs geometric mean overheads of 1.8\% and 9.9\% in client and server-side programs, respectively, while \codename-15ms introduces geometric mean overheads of 1.5\% and 8.8\% for these programs, respectively. 
%
%
The minor difference in performance between the two settings is not unexpected, even with a 1ms deadline \codename's performance is close to Native-OBS~(\cref{fig:overhead_realworld}). 
Therefore, in most cases, we believe that the cost of a 1ms protection deadline is affordable, especially given its security advantages~(\cref{subsec:case_study}).
%
%
Nevertheless, in performance-centric production server systems, \codename can be configured to a looser deadline (e.g., \textms{15}) to achieve better performance.

%

\section{Other Discussion}
\label{sec:discussion}
\PPn{Virtual machine support.}
On enterprise computers that want to run virtual machines, stage-2 page tables (S2PT) would be required by the hypervisor to isolate virtual machines.
In such systems, \codename can still leverage S2PT for log protection, if the security monitor interposes S2PT interfaces by using proposed techniques (\eg Nexen~\cite{Shi2017DeconstructingXen} and MyTEE~\cite{han2023mytee}).
Note that \codename's GPT-based implementation does not require such interposition, because GPT cannot be controlled at the hypervisor~(EL2) layer~\cite{rme-extension}. 

%

\PPn{TZASC-based isolation.}
%
%
TZASC~\cite{tzc400} is an alternative hardware protection primitive that can be applied.
The security monitor may maintain separated TZASC configurations for each protection domain (\cref{subsec:design_security_monitor}), which has been demonstrated by the prior work (i.e., Sanctuary~\cite{brasser2019sanctuary}).
However, TZASC controls memory permission at bus-transaction-level---all CPU cores within the cluster share the same permissions~\cite{cerdeira2022rezone}.
Unlike Sanctuary, \codename requires \hd to be frequently executed to consume and manage protected logs.
Whenever a CPU core executes within the protected domain (to execute \hd), all remaining cores in the cluster have to go idle, which may induce non-negligible runtime performance overhead.

\PPn{RISC-V and x86 architectural portability.}
%
%
On RISC-V, the security monitor can be implemented in the machine mode~\cite{lee2020keystone}, with Physical Memory Protection (PMP) for memory permission switching.
On x86, system management mode (SMM), such as UEFI firmware resource monitor or SMM-transfer monitor (STM), can be implemented as the monitor.
It configures the second-level address translation (\ie extended page table) and IOMMU to maintain protection domain isolation and permission switches~\cite{intel-stm,intel-frm}.
An alternative permission primitive is Protection Keys Supervisor (PKS~\cite{intel-pks}).
However, given the untrusted OS privilege can configure PKS interfaces, secure interposition is required.
Therefore, a security monitor follows Nested Kernel~\cite{daitenhahn2015nested} principles to interpose and protect all privileged instructions and isolate the PKS interface.
%
%



\PPn{Provenance-Assisted Denial of Service (PADoS) mitigation.}
Like for any system that enforces log protections~\cite{varun2023omnilog,ahmad2022hardlog}, although adversaries cannot compromise the log integrity and availability, they may perform the PADoS attack --- produce numerous logs in a short period to degrade system performance.
One potential solution is to adopt thread isolation for log processing~\cite{jiang2023auditing}.
Another approach is to reduce redundant logs and eliminate non-critical logs at generation~\cite{ma2018kcal,lim2021provbpf,sekar2024eaudit}.
Furthermore, the program and log profiling techniques can be adopted~\cite{kwon2018mci,PalanTir22} to train and filter benign behaviors.
We leave those optimizations as the future work.

\end{appendices}

\end{document}